\documentclass[12pt]{article}
\usepackage[utf8]{inputenc}
\usepackage{amsmath,amssymb,amsthm}
\usepackage{graphicx}
\graphicspath{{fig/}}
\usepackage[colorlinks=true,linkcolor=blue,citecolor=blue,filecolor=blue,urlcolor=blue,linktocpage=true]{hyperref}
\usepackage[numbers,sort&compress]{natbib} 
\usepackage{geometry}
\usepackage{authblk}
\usepackage[bottom]{footmisc}
\newcommand{\addEmail}[1]{\thanks{\href{mailto:#1}{#1}}}

\begin{document}

\title{Entanglement features in scattering mediated by heavy particles}
\author[1]{Chon Man Sou\addEmail{cmsou@mail.tsinghua.edu.cn}}
\author[2,3]{Yi Wang\addEmail{phyw@ust.hk}}
\author[2,3]{Xingkai Zhang\addEmail{xzhanghu@connect.ust.hk}}
\affil[1]{Department of Physics, Tsinghua University, Beijing 100084, China}
\affil[2]{Department of Physics, The Hong Kong University of Science and Technology,  Clear Water Bay, Kowloon, Hong Kong, P.R.China}
\affil[3]{The HKUST Jockey Club Institute for Advanced Study, The Hong Kong University of Science and Technology,  Clear Water Bay, Kowloon, Hong Kong, P.R.China}

\date{}
\maketitle

\begin{abstract}
    The amount of information propagated by an intermediate heavy particle exhibits characteristic features in inelastic scatterings with $n\geq 3$ final particles. As the total energy increases, the entanglement entropy, between its decay products and other final particles, exhibits a universal sharp dip, suppressed by its small decay rate. This indicates an entanglement suppression from a low-energy effective theory to a channel dominated by an on-shell heavy particle. As demonstrations of these entanglement features, we study concrete models of $2\to 3$ and $2\to 4$ scatterings, which shed light on the entanglement structure beyond the area law derived for $2\to 2$ scattering. In practice, these features may be probed by suitably marginalizing the phase-space distribution of final particles. 
\end{abstract}

\pagebreak

\tableofcontents
\pagebreak

\section{Introduction}

Entanglement is an important concept in high-energy physics with several well-known and
important applications, such as interpreting entanglement holographically \cite{Ryu:2006ef,Ryu:2006bv,Fursaev:2006ih}, describing conformal symmetry in CFT \cite{Calabrese:2009qy,Calabrese:2004eu}, and characterizing renormalization group (RG) flow \cite{Zamolodchikov:1986gt} by applying the entropic $c$-function \cite{Casini:2004bw,Casini:2006es}.

Recently, there has been growing interest in studying the concept of quantum entanglement in scattering and decaying processes at colliders, providing a natural scenario to explore entanglement in QFT experimentally and theoretically. Experimentally, various Bell tests utilizing spin states have been proposed, including top quark \cite{Afik:2020onf,Fabbrichesi:2021npl,Afik:2022kwm,Fabbrichesi:2022ovb} with observing the Bell violation from LHC data \cite{ATLAS:2023fsd}, gauge bosons \cite{Barr:2021zcp,Ashby-Pickering:2022umy,Fabbrichesi:2023cev,Barr:2022wyq,Aoude:2023hxv,Fabbrichesi:2024wcd}, lepton \cite{Altakach:2022ywa,Fabbrichesi:2022ovb,Ehataht:2023zzt,Ma:2023yvd} and photon pairs \cite{Fabbrichesi:2022ovb} (see \cite{Barr:2024djo} for a review of the setups), showing the feasibility to test quantum entanglement at high-energy scale. Theoretically, the entanglement in scattering and decaying processes has been studied by analyzing quantum-information quantities. These include entanglement entropies of various scatterings: $2\to 2$ \cite{Seki:2014cgq,Peschanski:2016hgk,Faleiro:2016lsf,Carney:2016tcs,Grignani:2016igg,Fan:2017hcd,Fan:2017mth,Peschanski:2019yah}, double scattering \cite{Fan:2021qfd}, cases with witness particles \cite{Araujo:2019mni,Fonseca:2021uhd,Shivashankara:2023koj} and deep inelastic scatterings \cite{Kharzeev:2017qzs}; entanglement entropy \cite{Lello:2013bva,Shivashankara:2023uvr}, concurrence \cite{Sakurai:2023nsc} and Mermin inequality \cite{Horodecki:2025tpn} for decaying processes, mutual information \cite{Fan:2017hcd,Araujo:2019mni,Fan:2021qfd}, relative entropy \cite{Bose:2020shm} and negativity \cite{Fedida:2022izl}. The implications of such quantum-information quantities have been explored in various aspects, including conditions for entropy growth \cite{Cheung:2023hkq,MacIntyre:2025cmr}, relation to QFT positivity \cite{Aoude:2024xpx}, the area law in scattering \cite{Fan:2017hcd,Low:2024mrk,Low:2024hvn}, relation to $S$-matrix bootstrap \cite{Bose:2020shm,Sinha:2022crx}, the flavor entanglement and its relations to symmetries \cite{Beane:2018oxh,Beane:2021zvo,Low:2021ufv,Liu:2022grf,Carena:2023vjc,Liu:2023bnr,Thaler:2024anb,Kowalska:2024kbs,McGinnis:2025brt} and phase transitions \cite{Liu:2025pny}. These studies suggest that quantum-information quantities may shed light on some properties of scattering and decaying from a novel perspective.

On the other hand, quantum-information quantities have been proposed to analyze effective field theory (EFT), such as the information loss by neglecting heavy fields \cite{Boyanovsky:2018fxl,Burgess:2024heo}, constraints by relative entropy between theories \cite{Cao:2022iqh,Cao:2022ajt} and perturbative unitary bounds \cite{Pueyo:2024twm,Cai:2025kcd}, providing a novel point of view to verify the validity of EFT. Since scattering processes are common scenarios to apply EFT, it might be inspiring to see how the breakdown of EFT manifests as features in quantum-information quantities in scattering. 

\begin{figure}
    \centering \includegraphics[width=0.5\linewidth]{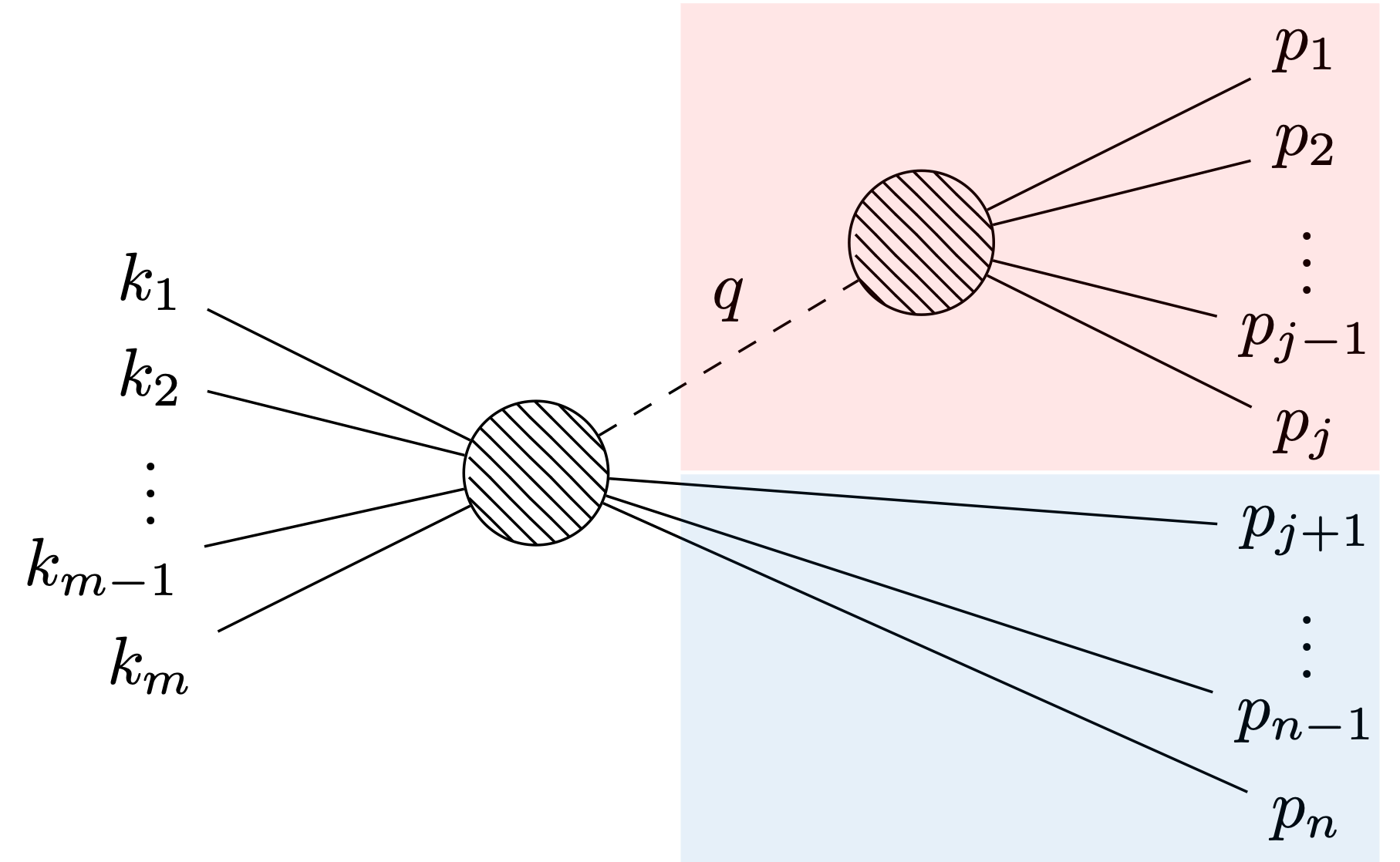}
    \caption{The information propagated by an intermediate heavy particle is quantified by the entanglement entropy between its decay products and other particles unrelated to the decay, highlighted in red and blue, respectively. For a realistic setup of $m\to n$ scattering, $j\geq 2$ and $n\geq j+1\geq 3$.}
    \label{fig:bipartitie_decay}
\end{figure}

In this work, we focus on the entanglement structure arising from intermediate heavy particles in scattering, with entanglement entropy in final state quantifying the amount of information they propagate. As shown in Figure~\ref{fig:bipartitie_decay}, we study the bipartite quantum system consisting of the decay products of the heavy particle and other particles unrelated to the decay. For the simple cases independent on any internal degree of freedom (e.g. spin, flavor and polarization), the two subsystems are described by two sets of momentum eigenstates $\{|p_1\rangle,\dots,|p_j\rangle\}$ and $\{|p_{j+1}\rangle,\dots,|p_n\rangle\}$ respectively, and the scattering amplitude describes the distribution of momentum configurations. The entanglement between the two subsystems is then governed by the heavy-field propagator
\begin{align}
    \frac{i}{q^2-M^2+i\Gamma M} \, , \label{eq:propagator_heavy}
\end{align}
with $q$, $M$ and $\Gamma$ as the momentum, mass and decay rate of the heavy particle, respectively, provided that $\Gamma\ll M$. As implied by the form of (\ref{eq:propagator_heavy}), in the low-energy regime, all the $q^2$ allowed by kinematics can evenly contribute to the entanglement, and the number of kinematically allowed configuration (multiplicity) scales with total energy, leading to a large amount of entanglement entropy scaling with total energy. On the other hand, when the total energy becomes  comparable and larger than $M$, the on-shell configuration with $q^2\approx M^2$ dominates the contribution, and this decreases the uncertainty of the effective region in phase space, leading to the reduction of entanglement entropy.\footnote{Since entropy measures the uncertainty of probabilistic mixture of subsystem state, if such a probability distribution is more localized in some region in phase space, the entropy decreases. As a pedagogical example, the differential entropy of the Cauchy distribution $\frac{1}{\pi}\frac{\Gamma M}{(z-M^2)^2+\Gamma^2 M^2}$ is $\log(4\pi \Gamma M)$, implying that if the width $\Gamma$ becomes smaller, the entropy decreases.} 

Following this intuitive picture, we perform a detailed analysis of the corresponding entanglement entropy, and key findings are highlighted as follows:
\begin{itemize}
    \item The on-shell contribution to the entanglement is universally suppressed by the small decay rate, and this suppression is attributed to the probabilistic mixture of the final quantum state of decay products by the heavy-field propagator (\ref{eq:propagator_heavy}).\footnote{We adopt the terminology used in the quantum-information textbook \cite{Nielsen:2012yss}. The {\it mixture of quantum state} means the density matrix with the form of linear combination $\rho=\sum_i p_i \rho_i$, where $p_i$ is a probability (distribution if the sum becomes a continuous integral), and $\rho_i$ are some density matrices. In this paper, the reduced density matrix of decay products, obtained by tracing out other final particles, has such a mixture form, such as (\ref{eq:rho_1_structure}) and (\ref{eq:rho_1j_mixture}) for $2\to 2$ and $m \to n$ scatterings respectively.}
    \item The entanglement in the low-energy regime is well described by the corresponding EFT, independent of and unsuppressed by the decay rate. This implies a dramatic reduction in entanglement entropy when the total energy is large enough for the on-shell limit $q^2\approx M^2$. 
\end{itemize}
These features are useful for distinguishing a complete theory from its low-energy EFT, and they are different from the area law commonly found in $2\to 2$ scatterings \cite{Fan:2017hcd,Low:2024mrk,Low:2024hvn}: the entropy typically scales with the total cross section.

The paper is organized as follows. In Section~\ref{sec:2-2_scattering}, we revisit the momentum-space entanglement of  both elastic and inelastic $2\to 2$ scatterings, with comment that the corresponding entropies have {\bf not} fully utilized the entanglement structure. In Section~\ref{sec:entanglement-mton}, we then explore the entanglement structure by a heavy particle in general $m\to n$ inelastic scattering, and the key features are derived with the on-shell approximation utilizing the properties of Cauchy distribution. In Section~\ref{sec:example}, we verify these features numerically in models of $2\to 3$ and $2\to 4$ scatterings, respectively. Section~\ref{sec:conclusion} is our conclusion.

\subsection*{Notation and conventions}
We work with the signature of $(+---)$. The four-momenta of initial and final particles are $k_i$ ($i=1,\dots,m$) and $p_i$ ($i=1,\dots,n$), respectively. To avoid unnecessary confusion, we denote the $i$-th particle's three-momentum by $\vec{p}_i$, and its magnitude is denoted by $\bar{p}_i=|\vec{p}_i|$. Its energy is determined by the on-shell condition $E_i=\sqrt{m_i^2+\bar{p}_i^2}$, and the norm of four-momentum is $p^2=(p^0)^2-\bar{p}^2$. Similarly, we use $\vec{k}_i$, $\bar{k}_i$ and $E_{k_i}$ for the initial particles.

For simplicity, we define the momenta $k_i$ and $p_i$ in the center-of-mass (CM) frame, so the total four-momentum
\begin{align}
    K=\sum_{i=1}^m k_i=\left(E_t,\vec{0}\right) \, ,
\end{align}
where $E_t$ is the total energy. With this, the $n$-body Lorentz-invariant phase space is denoted by
\begin{align}
    \int d\Pi_n(K;p_1,\dots,p_n) = \int\left(\prod_{i=1}^n\frac{d^3p_i}{(2\pi)^3 2E_i}\right) (2\pi)^4\delta^4\left(K-\sum_{i=1}^n p\right) \, ,
\end{align}
which will also be simply denoted by $\int d\Pi_n$ if neglecting the arguments does not cause any confusion.

The normalization of momentum eigenstate $|p\rangle$ with energy $E$ is
\begin{align}
    \langle p|p''\rangle=(2\pi)^32E\delta^3(\vec{p}-\vec{p}^{\,\prime\prime}) \, ,
\end{align}
where the volume of the system in the CM frame is
\begin{align}
    V=(2\pi)^3\delta^3(\vec{0}) \, ,
\end{align}
and similarly the interaction time is related to the delta function of energy
\begin{align}
    T=2\pi \delta(0) \, .
\end{align}
With these, we introduce another normalized eigenstate 
\begin{align}
    |\tilde{p}\rangle=\frac{1}{\sqrt{2EV}}|p\rangle \, ,
\end{align} 
with $\langle \tilde{p}|\tilde{p}\rangle=1$, which is more convenient while calculating density matrix and entanglement entropy.

\section{Revisit the momentum-space entanglement of $2\to 2$ scattering} \label{sec:2-2_scattering}
$2\to 2$ scattering is the simplest case to study entanglement between final particles. The information propagated during such an scattering simply satisfies an area law, depending on the total cross section, and this has been shown with entanglement entropy \cite{Seki:2014cgq,Fan:2017hcd} and other forms of entropy \cite{Low:2024mrk,Low:2024hvn}. 

In Section~\ref{sec:2-2setup}, we first revisit the setup for calculating the entanglement entropy $S_{EE}^{(2\to 2)}$ in $2\to 2$ scatterings independent of any internal degree of freedom, with emphasis on the contribution dominated by an on-shell heavy particle. In Section~\ref{sec:SEE22_structure}, we discuss the structure of such $S_{EE}^{(2\to 2)}$ for both elastic and inelastic scatterings, where the information propagated by on-shell heavy particles is proportional to the total cross section in the elastic case. We also emphasize the logarithmic factor associated with the multiplicity of kinematically allowed final state, often overlooked in the literature, whereas this is not only important in the inelastic case, but also in decaying processes \cite{Lello:2013bva} and the entanglement structure beyond the area law discussed in Section~\ref{sec:entanglement-mton}.

\subsection{Setup of $2\to 2$ scattering}\label{sec:2-2setup}
We consider the initial state as an unentangled two-particle state $| i \rangle = \frac{1}{\sqrt{4E_{k_1}E_{k_2} }V} | k_1,k_2\rangle$.
It evolves into the final state related to the $S$-matrix by inserting a complete basis
\begin{align}
| f \rangle &= \hat{S} |i\rangle \nonumber \\
&= \frac{1}{\sqrt{4E_{k_1}E_{k_2}}V} \sum_n \int \left(\prod_{i=1}^n\frac{d^3p_i}{(2\pi)^32E_i}\right) | p_1, \dots, p_n\rangle\langle p_1, \dots, p_n | \hat{S} | k_1,k_2\rangle \,,
\end{align}
where we assume that all particles are distinguishable throughout this paper, by considering different types of particles. Suppose we only care about the $2\rightarrow 2$ scattering consisting of particles $1$ and $2$ by inserting only the corresponding two-particle states, other parts with $n\neq 2$ are neglected at the expense of introducing a probability of the selected $2\rightarrow 2$ scattering occurring: $P^{(2\rightarrow 2)}$ \cite{Low:2024hvn}. Now we can write
\begin{align}
| f^{(2\rightarrow 2)} \rangle &= \frac{1}{\sqrt{P^{(2\rightarrow 2)}}}\frac{1}{\sqrt{4E_{k_1}E_{k_2} }V} \int \frac{d^3p_1}{(2\pi)^32E_1} \frac{d^3p_2}{(2\pi)^32E_2} \langle p_1,p_2 | \hat{S} | k_1,k_2\rangle | p_1,p_2\rangle\,, \label{eq:f2to2_project}
\end{align}
where $P^{(2\to 2)}$ also serves to normalize the final state $| f^{(2\to 2)}\rangle$, satisfying $\langle f^{(2\rightarrow 2)} | f^{(2\rightarrow 2)}\rangle = 1$.
The matrix element is
\begin{align}
&\langle p_1,p_2 | \hat{S} | k_1,k_2\rangle \nonumber \\
=& \langle p_1,p_2 | 1+i \hat{T} | k_1,k_2\rangle \nonumber \\
=& \langle p_1,p_2 | k_1,k_2\rangle + i \mathcal{M} (2 \to 2) (2\pi)^4\delta^4(K-p_1-p_2) \nonumber \\
=& \delta_{\rm el}4E_{k_1}E_{k_2}(2\pi)^6\delta^3(\vec{p}_1-\vec{k}_1)\delta^3(\vec{p}_2-\vec{k}_2) + i \mathcal{M} (2 \to 2) (2\pi)^4\delta^4(K-p_1-p_2)  \,, \label{eq:2-2-S-matrix}
\end{align}
where the prefactor of the first term survives for elastic scattering (the selected two final particles same as the initial two) and vanishes otherwise
\begin{align}
    \delta_{\rm el}=\begin{cases}
        1 \, , & {\rm elastic}\\
        0 \, , & {\rm inelastic}
    \end{cases} \, .
\end{align}
With (\ref{eq:f2to2_project}) and (\ref{eq:2-2-S-matrix}), the $2\to 2$ final state becomes
\begin{align}
| f^{(2\rightarrow 2)} \rangle = \frac{1}{\sqrt{P^{(2\rightarrow 2)}}}\frac{1}{\sqrt{4E_{k_1}E_{k_2}}V} \left(\delta_{\rm el}| k_1,k_2\rangle+\int d\Pi_2\, i\mathcal{M}(2\to 2)|p_1,p_2\rangle\right) \, ,
\end{align}

For the density matrix of the final pure state $\rho^{(2\rightarrow 2)} = | f^{(2\rightarrow 2)}\rangle\langle f^{(2\rightarrow 2)} |$,  the reduced density matrix of particle $1$ is obtained by tracing out particle $2$:
\begin{align}
\rho^{(2\rightarrow 2)}_1 &= \int \frac{d^3p_2}{2E_2(2\pi)^3}\langle p_2 | \rho^{(2\rightarrow 2)} | p_2\rangle \nonumber \\
&= \frac{\delta_{\rm el}}{P^{(2\rightarrow 2)}}\Bigg[1 - \Delta \sum_n \int d\Pi_n | \mathcal{M} (2 \to n) |^2\Bigg] | \tilde{k}_1\rangle\langle \tilde{k}_1 | \nonumber \\
&\quad + \frac{\Delta}{P^{(2\rightarrow 2)}}\int d\Pi_2| \mathcal{M} (2 \to 2) |^2 | \tilde{p}_1\rangle\langle \tilde{p}_1 | \, , \label{eq:rho1_22_F_NF}
\end{align}
where the optical theorem
\begin{equation}
2{\rm Im} \mathcal{M} (2 \to 2) = \sum_n \int d\Pi_n | \mathcal{M} (2 \to n) |^2 \,,
\end{equation}
is applied for the elastic case, and we introduce a dimensionless parameter for initial momenta
\begin{align}
    \Delta &= \frac{(2\pi)^4\delta^4(K-p_1-p_2)}{4E_{k_1}E_{k_2}V^2} \nonumber \\
    &=\frac{2\pi \delta(E_t-E_1-E_2)}{4E_{k_1}E_{k_2}V} \nonumber \\
    &= \frac{T}{4E_{k_1}E_{k_2}V}
\end{align} with $T$ as the interaction time in the CM frame.\footnote{Following the textbook's treatment \cite{Weinberg:1995mt}, we can consider the system is in a time box, and the delta function of energy now becomes $2\pi\delta_T(E_t-E_1-E_2)=\int^{\frac{T}{2}}_{-\frac{T}{2}}dt\, e^{i(E_t-E_1-E_2)t}$, 
which clearly implies that $2\pi \delta(0)=T$ when $T$ is large and the energy conservation $E_t=E_1+E_2$ satisfies for the $2\to 2$ scattering. } The reduced density matrix (\ref{eq:rho1_22_F_NF}) consists of two terms: the first corresponds to the forward scattering, and the second term originates from the scattered part.  $\Delta$ can be justified as a small quantity for perturbative expansion, by a detailed proof using wave packets in \cite{Kowalska:2024kbs}.

After having the reduced density matrix, we are able to calculate the entanglement entropy. In order to calculate the entanglement entropy, we first discretize the momentum and 2-body phase spaces:

\begin{align}
\label{eq:discretize_dPi2_aligned}
V\int \frac{d^{3}p}{(2\pi)^{3}} &= \sum_{p} \\
\int\!d\Pi_2 &= \int \frac{d^{3}p_1}{(2\pi)^{3}2E_1}\frac{d^{3}p_2}{(2\pi)^{3}2E_2}(2\pi)^{4}\delta^4(K-p_1-p_2) \nonumber \\
&= \int \frac{d^{3}p_1}{(2\pi)^{3}}\frac{2\pi \delta(E_t-E_1-E_2)}{2E_1 2E_2} \nonumber \\
&= \sum_{p_1} \Delta_{p_1} \,.
\end{align}
where $E_1$, $E_2$ are on-shell and fixed by the conservation of energy, and 
\begin{equation}
\Delta_{p_1}=\frac{2\pi\delta(E_t-E_1-E_2)}{4E_1E_2V} \, , \label{eq:Deltap}
\end{equation}
is a dimensionless parameter for a given final momentum $\vec{p}_1$, which is useful to determine the multiplicity of final states allowed by kinematics.

Notice that the reduced density matrix (\ref{eq:rho1_22_F_NF}) is diagonal, so the entanglement entropy between the two final particles can be calculated easily
\begin{align}\label{eq:SEE}
S^{(2\rightarrow 2)}_{EE} &= -\mathrm{Tr}\left(\rho^{(2\rightarrow 2)}_1\log\rho^{(2\rightarrow 2)}_1\right) \nonumber \\
&= -\frac{P_F}{P^{(2\rightarrow 2)}}\log\frac{P_F}{P^{(2\rightarrow 2)}}\nonumber \\
&\quad- \sum_{p_1} \frac{\Delta\Delta_{p_1}}{P^{(2\rightarrow 2)}} |\mathcal{M}(2\to 2)|^2  \log\left[\frac{\Delta\Delta_{p_1}}{P^{(2\rightarrow 2)}} |\mathcal{M}(2\to 2)|^2\right] \nonumber \\
&= -\frac{P_F}{P^{(2\rightarrow 2)}}\log\frac{P_F}{P^{(2\rightarrow 2)}}\nonumber \\
&\quad-\int d\Pi_2 \frac{\Delta}{P^{(2\rightarrow 2)}} |\mathcal{M}(2\to 2)|^2  \log\left[\frac{\Delta\Delta_{p_1}}{P^{(2\rightarrow 2)}} |\mathcal{M}(2\to 2)|^2\right]\,,
\end{align}
where 
\begin{align}
    P_F=\delta_{\rm el}\left(1 - \Delta \sum_n \int d\Pi_n | \mathcal{M} (2 \to n) |^2 \right)\, ,
\end{align}
is the probability of forward scattering. We can recognize that the probability density for the initial state being scattered into the component with momenta $p_1,p_2$ (where $p_2=K-p_1$) is given by:
\begin{equation}\label{eq:prob}
\Delta\Delta_{p_1} |\mathcal{M}(2\to 2)|^2\,,
\end{equation}
so the probability of not being forward scattering in the $2\to 2$ scattering process is
\begin{equation}\label{eq:PNF}
P_{NF} = \sum_{p_1} \Delta\Delta_{p_1} |\mathcal{M}(2\to 2)|^2 = \int d\Pi_2\, \Delta |\mathcal{M}(2\to 2)|^2 \, . 
\end{equation}
It is noteworthy that $P_F+P_{NF}=P^{(2\to 2)}$ by the the normalization of $\rho_1^{(2\to 2)}$ (\ref{eq:rho1_22_F_NF}):
\begin{align}
    1 =& {\rm Tr}\rho^{(2\to 2)}_1 \nonumber \\
=& \frac{\delta_{\rm el}}{P^{(2\to 2)}}\Bigg[1 - \Delta \sum_n \int d\Pi_n | \mathcal{M}(2\to n) |^2 \Bigg] + \frac{\Delta}{P^{(2\to 2)}}\int d\Pi_2 | \mathcal{M}(2\to 2) |^2 \nonumber \\
=& \frac{P_N+P_{NF}}{P^{(2\to 2)}}\,.
\end{align}

We are now ready to calculate the entanglement entropy between the two final particles. In the CM frame, the size of the momentum is determined by
\begin{align}
\sqrt{\bar{p}^{\,2}_{\text{CM}}+m_1^2} + \sqrt{\bar{p}^{\,2}_{\text{CM}}+m_2^2} &= \sqrt{s} \nonumber \\
\bar{p}_{CM}(\sqrt{s};m_1,m_2) &= \sqrt{\frac{(m_2^2-m_1^2)^2-2s(m_2^2+m_1^2)+s^2}{4s}} \,, 
\label{eq:CMframe_pCM}
\end{align}
where $s=E_t^2$ is one of the Mandelstam variables, and this form of CM momentum will appear in many results of the paper. This determines the state in phase space allowed by the energy-momentum conservation:
\begin{align}
2\pi\delta(E_1+E_2-\sqrt{s}) &= 2\pi\frac{E_1 E_2}{\bar{p}_{ CM}\sqrt{s}}\delta(\bar{p}_1-\bar{p}_{CM}) \,.
\label{eq:deltaE_final}
\end{align}
If the scattering amplitude is independent of direction, we can obtain a simple analytical expression, as the second part of (\ref{eq:SEE}) is proportional to a $2$-body phase-space integral calculable with (\ref{eq:deltaE_final}):
\begin{align}
&-\int d\Pi_2 \frac{\Delta}{P^{(2\rightarrow 2)}} |\mathcal{M}(2\to 2)|^2  \log\left[\frac{\Delta\Delta_{p_1}}{P^{(2\rightarrow 2)}} |\mathcal{M}(2\to 2)|^2\right] \nonumber \\
=& -\int\frac{d^3p_1}{(2\pi)^2}\frac{\Delta}{P^{(2\rightarrow 2)}}\frac{\delta(\bar{p}_1-\bar{p}_{CM})}{4\bar{p}_{ CM}\sqrt{s}} |\mathcal{M}(2\to 2)|^2 \log\left(\frac{\Delta}{P^{(2\to 2)}}|\mathcal{M}(2\to 2)|^2 \frac{ T}{4E_1E_2V}\right)\nonumber \\
=& -\frac{\Delta }{P^{(2\rightarrow 2)}} \frac{\bar{p}_{CM}}{4\pi\sqrt{s}}|\mathcal{M}(2\to 2)|^2 \log\left(\frac{\Delta}{P^{(2\to 2)}}|\mathcal{M}(2\to 2)|^2 \frac{ T}{4E_1E_2V}\right)\,,
\label{eq:SEE_full_derivation}
\end{align}
where (\ref{eq:Deltap}) is applied in the logarithm, and the energy delta function $2\pi \delta(E_t-E_1-E_2)$ is evaluated as $T$. Combining with (\ref{eq:SEE}) gives the entanglement entropy:
\begin{align}\label{eq:SEE full}
&S_{EE}^{(2\to 2)}  \nonumber \\
=& -\frac{P_F}{P^{(2\rightarrow 2)}}\log\frac{P_F}{P^{(2\rightarrow 2)}}-\frac{\Delta }{P^{(2\rightarrow 2)}} \frac{\bar{p}_{CM}}{4\pi\sqrt{s}}|\mathcal{M}(2\to 2)|^2 \log\left(\frac{\Delta}{P^{(2\to 2)}}|\mathcal{M}(2\to 2)|^2 \frac{ T}{4E_1E_2V}\right) 
\, .
\end{align}

Clearly, among the three channels of scattering amplitudes ($s,t,u$), only the $s$-channel is isotropic. In the on-shell limit of the heavy particle, $s\approx M^2$, the $s$-channel amplitude dominates over other channels, so that we can neglect the contributions of other channels.

\subsection{The structure of $S_{EE}^{(2\to 2)}$ and the area law}
\label{sec:SEE22_structure}
By using the definition of $P_{NF}$ (\ref{eq:PNF}) for the isotropic $\mathcal{M}(2\to 2)$, the entanglement entropy (\ref{eq:SEE full}) can be rewritten as 
\begin{align}
    S_{EE}^{(2\to 2)}&=-\frac{P_F}{P^{(2\rightarrow 2)}}\log\frac{P_F}{P^{(2\rightarrow 2)}}-\frac{P_{NF}}{P^{(2\rightarrow 2)}}\log\frac{P_{NF}}{P^{(2\rightarrow 2)}}+\frac{P_{NF}}{P^{(2\to 2)}}\log\left(\frac{\bar{p}_{CM}E_1E_2V}{\pi\sqrt{s}T}\right) \nonumber \\
    &= 
    \begin{cases}
        -\frac{P_{NF}}{P^{(2\rightarrow 2)}}\log\frac{P_{NF}}{P^{(2\rightarrow 2)}} +\mathcal{O}(\Delta) \, , & {\rm elastic} \\
        \log\left(\frac{\bar{p}_{CM}E_1E_2V}{\pi\sqrt{s}T}\right)  \, , & {\rm inelastic}
    \end{cases} \, , \label{eq:SEE22_structure}
\end{align}
where we applied the fact that $P_{NF}\approx\mathcal{O}(\Delta)$ in (\ref{eq:PNF}). As $P_{NF}$ is proportional to the total cross section $\sigma_{\rm tot}$ in the elastic case, this indicates an area law that $S_{EE}^{(2\to 2)}$ grows with $\sigma_{\rm tot}$ \cite{Seki:2014cgq,Fan:2017hcd}. On the other hand, $S_{EE}^{(2\to 2)}$ of the inelastic case is independent of the existence of heavy particle, but depends on quantities fixed by kinematics.

Things become less mysterious if we understand the structure of the reduced density matrix (\ref{eq:rho1_22_F_NF}), and it is essentially a mixture of states of forward and non-forward scatterings, meaning that it is in a form of linear combination
\begin{align}
    \rho_1^{(2\to 2)}=\sum_{a=F,NF} \frac{P_a}{P^{(2\to2)}}\rho_a \, , \label{eq:rho_1_structure}
\end{align}
where the forward part exists for elastic scattering $\rho_F=|\tilde{k}_1\rangle\langle\tilde{k}_1|$, and the non-forward part describes the multiplicity of final states allowed by kinematics
\begin{align}
    \rho_{NF}&=\frac{\int d\Pi_2|\mathcal{M}(2\to2)|^2|\tilde{p}_1\rangle \langle \tilde{p}_1|}{\int d\Pi_2|\mathcal{M}(2\to2)|^2} \nonumber \\
    &=\sum_{p_1}\frac{\pi\sqrt{s}T}{\bar{p}_{CM}E_1E_2V} |\tilde{p}_1\rangle \langle \tilde{p}_1| \, , 
\end{align}
for isotropic $\mathcal{M}(2\to 2)$, where $\int d\Pi_2=\frac{\bar{p}_{CM}}{4\pi\sqrt{s}}$ and its discretization (\ref{eq:Deltap}) are applied. Since $\rho_F$ and $\rho_{NF}$ are diagonal and the associated subspaces are orthogonal, the von Neumann entropy of (\ref{eq:rho_1_structure}) is well-known in textbook \cite{Nielsen:2012yss} that
\begin{align}
    S\left(\rho_1^{(2\to 2)}\right)=H\left(\frac{P_a}{P^{(2\to 2)}}\right)+\sum_{a=F,NF}\frac{P_a}{P^{(2\to 2)}}S(\rho_a) \, , \label{eq:S_mixture}
\end{align}
where $H(P)$ is the classical Shannon entropy, and this clearly separates the contributions from mixture (with $P_a/P^{(2\to 2)}$) and quantum states (with $\rho_a$) in (\ref{eq:SEE22_structure}). When entropic structures are discussed in this paper, the {\it mixture part} indicates the contribution of the Shannon entropy $H(p_a)$, and the {\it state's part} means the part related to the microscopic detail of density matrices $\rho_a$, such as the multiplicity allowed by kinematics.

With the intuitive meaning of (\ref{eq:S_mixture}), the second line of (\ref{eq:SEE22_structure}) indicates the following facts:
\begin{itemize}
    \item For elastic scattering, only the mixture part depending on the total cross section dominates the contribution to $S_{EE}$, leading to the area law.
    \item For inelastic scattering, $S_{EE}$ only depends on the multiplicity of the kinematically allowed states, not including any quantity related to the intermediate heavy particle.
\end{itemize}
Therefore, the entanglement entropy of $2\to 2$ scattering has {\bf not} fully utilized the entanglement structure. In particular, if there exist entanglement features depending on both the properties of an intermediate heavy particle and the multiplicity by kinematics, this immediately implies going beyond the $2 \to 2$ area law.\footnote{To be precise, we mean that the features still exist in some limits for discretization factor like $\Delta \to 0$, such as the results shown in (\ref{eq:SEE22_structure}).}

Before ending the section of $2\to 2$ scattering, we also comment the above reasoning with other forms of entropies, such as the $n$-Tsallis and $n$-R\'enyi with integer $n\geq 2$. For convenience, we express the diagonal $\rho_F$ and $\rho_{NF}$ as
\begin{align}
    \rho_a=\sum_b q_{ab}|\tilde{\psi}_{ab}\rangle\langle\tilde{\psi}_{ab}| \, ,
\end{align}
with a suitable orthonormal basis, and we have
\begin{align}
    {\rm Tr}\left(\rho_1^{(2\to 2)}\right)^n&=\sum_{a,b}\left(\frac{P_a}{P^{(2\to 2)}}q_{ab}\right)^n \nonumber \\
    &=\sum_a\left(\frac{P_a}{P^{(2\to 2)}}\right)^n+\sum_a\left(\frac{P_a}{P^{(2\to 2)}}\right)^n\left(\sum_bq_{ab}^n-1\right) \, .
\end{align}
With this, the $n$-Tsallis quantum entropy of $\rho_1^{(2\to 2)}$ is
\begin{align}
    S_{n,T}\left(\rho_1^{(2\to 2)}\right)&=\frac{1-{\rm Tr}\left(\rho_1^{(2\to 2)}\right)^n}{n-1} \nonumber \\
    &=H_{n,T}\left(\frac{P_a}{P^{(2\to 2)}}\right)+\sum_a\left(\frac{P_a}{P^{(2\to 2)}}\right)^nS_{n,T}\left(\rho_a\right) \nonumber \\
    &=\begin{cases}
        \frac{n}{n-1}\frac{P_{NF}}{P^{(2\to 2)}}+\mathcal{O}(\Delta^2) \, , & {\rm elastic} \\
        \frac{1-\left(\frac{\pi\sqrt{s}T}{\bar{p}_{CM}E_1E_2V}\right)^{n-1}}{n-1}\approx\frac{1}{n-1}+\mathcal{O}(\Delta_{p_1}^{n-1}) \, , & {\rm inelastic}
    \end{cases}\, , \label{eq:Tsallis_22}
\end{align}
where $H_{n,T}$ is the $n$-Tsallis classical entropy of $\frac{P_a}{P^{(2\to 2)}}$. Similar to (\ref{eq:SEE22_structure}), the cases of elastic and inelastic scatterings are contributed by mixture and state's multiplicity respectively, but there are some differences: the former is now mainly contributed by the forward term with $P_F/P^{(2\to 2)}=1-P_{NF}/P^{(2\to 2)}$, and the latter does not scale with the large multiplicity. On the other hand, the $n$-R\'enyi entropy is
\begin{align}
    S_{n,R}\left(\rho_1^{(2\to 2)}\right)&=\frac{1}{1-n}\log\left({\rm Tr}\left(\rho_1^{(2\to 2)}\right)^n\right) \nonumber \\
    &=\frac{1}{1-n}\log\left(1-(n-1)\left[H_{n,T}\left(\frac{P_a}{P^{(2\to 2)}}\right)+\sum_a\left(\frac{P_a}{P^{(2\to 2)}}\right)^nS_{n,T}\left(\rho_a\right)\right]\right) \nonumber \\
    &=\begin{cases}
        \frac{n}{n-1}\frac{P_{NF}}{P^{(2\to 2)}}+\mathcal{O}(\Delta^2) \, , & {\rm elastic} \\
         \log\left(\frac{\bar{p}_{CM}E_1E_2V}{\pi\sqrt{s}T}\right) \, , & {\rm inelastic}
    \end{cases} \, ,
\end{align}
where the elastic case agrees with the $n$-Tsallis and indicates an area law, as demonstrated in \cite{Low:2024hvn}, whereas the inelastic case agrees with the scaling of multiplicity as $S_{EE}$. Therefore, while using other forms of entropies, it is possible to neglect some entanglement structures in perturbation theory, and the conclusion still holds as $S_{EE}$ that the $S_{n,T}$ and $S_{n,R}$ of $2\to 2$ scattering have not fully utilized the entanglement structure.

\section{The entanglement structure from heavy particle in $m\to n$ scattering}\label{sec:entanglement-mton}
In particle physics, it is known that a decaying process can generate entanglement between final particles, yet its entropy does not follow the area law. As demonstrated in \cite{Lello:2013bva} with the time-dependent Wigner-Weisskopf method, the entanglement entropy for a heavy particle with mass $M$ decaying into two particles $1$ and $2$ is
\begin{align}
    S_{EE}^{(1\to 2)}=\log\left(\Gamma\frac{\bar{p}_{CM}E_1E_2V}{\pi M }\right) \, , \label{eq:SEE_decay}
\end{align}
where all the quantities are defined in the CM frame.\footnote{It is that noteworthy there is a factor of $4$ difference in the logarithm compared to \cite{Lello:2013bva}, and we will show in Section~\ref{sec:on-shell_Cauchy} that it is attributed to a mistake of applying a ``sharp-peak limit" to calculate entropy.} Clearly, (\ref{eq:SEE_decay}) does not include any proportionality constant as in $2\to 2$ elastic scattering, whereas it has the form related to the multiplicity in phase space, similar to the inelastic scattering (\ref{eq:SEE22_structure}) by replacing the interaction time $T$ to the lifetime $\Gamma^{-1}$. Generally, such a decaying process is included in a larger scattering process where the initial and final asymptotic states are well-defined. Our goal is to identify the entanglement related to decay and how it varies in general $m\to n$ scattering, which is expected to go beyond the $2 \to 2$ area law.

\begin{figure}
	\centering
	\includegraphics[width=\textwidth]{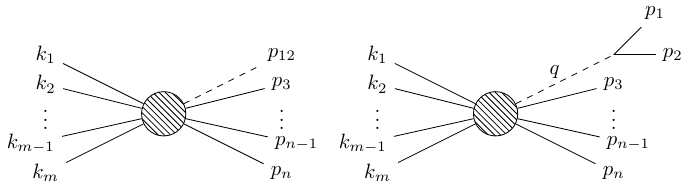}
	\caption{{\bf Left:} A general inelastic $m\to n-1$ scattering, represented by the part $i\mathcal{M}(m \to n-1)$, in which a heavy particle with momentum $p_{12}$ is produced. Solid and dashed lines represent light and heavy particles respectively, in the sense that it is heavy enough to decay into two light particles. {\bf Right:} The heavy particle further decays into two light particles with momenta $p_1$ and $p_2$ respectively, leading to a $m \to n$ scattering, represented by $i\mathcal{M}(m \to n)$. The decay products generally entangle with the rest of the particles with momenta $p_3,\, \dots\, , \, p_n$. \label{fig:general_n_decay_12}}
\end{figure}

Our approach is to approximate the calculation of entanglement entropy of general $m\to n$ scattering involving a heavy particle, by analyzing the pole structure of the amplitudes. As a concrete example, the two amplitudes in Figure~\ref{fig:general_n_decay_12} are related by the Breit-Wigner formula near the resonance
\begin{align}
	i\mathcal{M}(m\to n)&\approx i\mathcal{M}(m\to n-1) \frac{g}{(p_1+p_2)^2-M^2+iM\Gamma} \, , \label{eq:amplitude_n_n-1}
\end{align}
where the propagator of heavy particle has a sharp peak at the on-shell condition $(p_1+p_2)^2=M^2$ when $\Gamma\ll M$, as its absolute square has the form of the Cauchy distribution $\mathcal{P}_{\Gamma,M}(p^2)$:
\begin{align}
	\frac{g^2}{\left[(p_1+p_2)^2-M^2\right]^2+M^2\Gamma^2} &=\frac{\pi g^2}{\Gamma M}\left\{\frac{1}{\pi}\frac{\Gamma M}{\left[(p_1+p_2)^2-M^2\right]^2+\Gamma^2M^2}\right\} \nonumber \\
	&=\frac{8\pi^2 M}{\bar{p}_{CM}} \mathcal{P}_{\Gamma,M}((p_1+p_2)^2)\, , \label{eq:propagator_Cauchy}
\end{align}
where the formula for the $1\to 2$ decay rate $\Gamma$ is applied, and the CM momentum is $\bar{p}_{CM}(M;m_1,m_2)$ defined in (\ref{eq:CMframe_pCM}).

In the following subsections, we will utilize the Cauchy distribution appearing in (\ref{eq:propagator_Cauchy}) to derive the on-shell approximation of entanglement entropy. In Section~\ref{sec:on-shell_Cauchy}, we use the complex analysis to derive approximations for one-dimensional integrals and entropies involving the Cauchy distribution in the small $\Gamma$ limit. In Section~\ref{sec:n-particle-recursion}, we apply the phase-space recursive relation to decompose $n$-body integrals, in which we derive the decomposition of phase-space integrals and exact form of reduced density matrix of general $j$ decay particles. Section~\ref{sec:structure_decay_onshell} includes the key results: we derive a general entanglement structure between decay products and other final particles, highlighting that the heavy-field propagator contributes to the entanglement entropy in the form of probabilistic mixture of quantum states, as shown in (\ref{eq:SEE_1j_mixture}).\footnote{In this case, the reduced density matrix of decay products is a mixture of quantum states: $\rho_R=\int dq^2 \, P(q^2)\sigma(q^2)$, a linear combination over different $q^2$.} Based on this fact, we derive an approximation describing the amount of entropy attributed to the on-shell heavy particle, provided that the decay rate is small compared to its mass, as shown in (\ref{eq:SEE_1j_on-shell}). In Section~\ref{sec:jn-1}, we demonstrate a concrete case that the decay products entangling with only one particle, where subtle dependence on spacetime volume is discussed. We also comment that the entanglement features based on the reduced density matrix of decay products can in principle be probed by suitably marginalizing the phase-space distribution of final particles.

\subsection{The on-shell approximation with the Cauchy distribution}\label{sec:on-shell_Cauchy}
Firstly, we demonstrate how one-dimensional integrals with the Cauchy distribution is approximated in the case with small $\Gamma M$, suppressed by $g^2$. Suppose the function to be integrated satisfies
\begin{itemize}
	\item $|f(z)|\lesssim\mathcal{O}(z)$ for $z\to \infty$. This agrees with the kinematics fixed by finite total energy.
	\item Its simple poles $z_i$ on the upper-half plane are separated from the poles of the Cauchy distribution $\mathcal{P}_{\Gamma,M}(z)$, with separations not suppressed by the small $\Gamma M$. This is consistent to the requirement of applying the Breit-Wigner formula (\ref{eq:amplitude_n_n-1}).
	\item The end points of integration interval $[z_L,z_R]$ are separated from the poles of $\mathcal{P}_{\Gamma,M}(z)$, which is justified as follows. For general scatterings, the lower limit $z_L$ corresponds to the energy needed to produce outgoing light particles, much smaller than the invariant heavy mass $M^2$. The upper limit $z_R$ is controlled by the initial total energy $K^2=E^2_t$, sufficiently larger than the threshold of producing the on-shell heavy particle ($z_R-M^2\gg \Gamma M$), for making the on-shell approximation accurate.
\end{itemize}
With these conditions, we first calculate the integral over $(-\infty,+\infty)$, deforming into a contour integral of the upper-half complex plane
\begin{align}
	&\int^{+\infty}_{-\infty}dz\, f(z) \mathcal{P}_{\Gamma,M}(z)  \nonumber \\
	=& f(M^2+i\Gamma M)+2\pi i \sum_i  \frac{{\rm Res}(f,z_i)}{\pi}\frac{\Gamma M}{\left(z_i-M^2\right)^2+\Gamma^2M^2} -i\pi \lim_{z\to \infty}\frac{f(z)}{\pi z} \Gamma M  \nonumber \\
	\approx& f(M^2)+\left[if'(M^2)+2i\sum_{i}\frac{{\rm Res}(f,z_i)}{(z_i-M^2)^2}-i \lim_{z\to \infty}\frac{f(z)}{ z}\right]\Gamma M +\mathcal{O}\left(\frac{\Gamma^2}{M^2}\right)
	\, , \label{eq:Cauchy_expectation}
\end{align}
where we use the fact that 
\begin{align}
	2\pi i {\rm Res}(\mathcal{P}_{\Gamma,M},M^2\pm i \Gamma M)=\pm 1 \, .
\end{align}
By an order estimation, we know that the contribution by $z\notin[z_L,z_R]$ is suppressed by small $\Gamma M$:
\begin{align}
	\mathcal{P}_{\Gamma,M}(z)
	&=\frac{1}{\pi}\frac{\Gamma M}{(z-M^2)^2+\Gamma^2 M^2}  \nonumber \\
	&\propto \Gamma M  \, ,
\end{align}
implying that 
\begin{align}
	\int^{z_R}_{z_L}dz\, f(z) \mathcal{P}_{\Gamma,M}(z) \approx \int^{+\infty}_{-\infty}dz\, f(z) \mathcal{P}_{\Gamma,M}(z) +\mathcal{O}\left(\frac{\Gamma} {M}\right) \, .
\end{align}
In the sense of (\ref{eq:Cauchy_expectation}), the Cauchy distribution can be approximated by a Dirac delta function
\begin{align}
	\mathcal{P}_{\Gamma,M}(z) &\approx \delta\left(z^2-M^2\right) \, , \label{eq:on-shell_approx}
\end{align}
and the correction to the delta function, suppressed by small $\Gamma M$, can be justified by the contour integral perturbatively.

On the other hand, we need a more careful approximation for calculating entropy with the Cauchy distribution as follows. Now we suppose the following stronger conditions for $f(z)$: 
\begin{itemize}
	\item $|f(z)|<\mathcal{O}(z)$ for $z\to \infty$. This again agrees with the kinematics fixed by finite total energy.
	\item Its poles $z_i$ (not on the real axis) and the corresponding branch cut of $f(z)\log(f(z))$ are separated from the poles of the Cauchy distribution $\mathcal{P}_{\Gamma,M}(z)$, with separations not suppressed by the small $\Gamma M$. This is again consistent to the requirement of applying the Breit-Wigner formula (\ref{eq:amplitude_n_n-1}).
	\item The end points of integration interval $[z_L,z_R]$ are separated from the poles of $\mathcal{P}_{\Gamma,M}(z)$, same as the previous condition.
\end{itemize}
With these conditions, for a diagonal density matrix\footnote{Note that we are not calculating the differential entropy $-\int dz\, f(z) \mathcal{P}_{\Gamma,M}(z) \log\left( f(z) \mathcal{P}_{\Gamma,M}(z)\right)$, which does not involve the discretization factor $\Delta_z$.}
\begin{align}
	\rho&=\int dz \, f(z)\mathcal{P}_{\Gamma,M}(z)|\tilde{z}\rangle \langle \tilde{z}| \nonumber \\
	&=\sum_z \Delta_z f(z)\mathcal{P}_{\Gamma,M}(z)|\tilde{z}\rangle \langle \tilde{z}| \, ,
\end{align}
where $\Delta_z$ discretizes the measure $dz$ and the orthonormal basis satisfies $\langle \tilde{z}|\tilde{z}\rangle=1$, its entropy can be approximated as follows
\begin{align}
	&-\int^{+\infty}_{-\infty}dz\, f(z) \mathcal{P}_{\Gamma,M}(z) \log\left(\Delta_z f(z) \mathcal{P}_{\Gamma,M}(z)\right) \nonumber \\
	=&-\int^{+\infty}_{-\infty}dz\, f(z) \mathcal{P}_{\Gamma,M}(z) \Bigg[\log\left(\sqrt{\frac{\Delta_z \Gamma M}{\pi}}\frac{1}{z-M^2+i\Gamma M} \right) \nonumber \\
	&+\log \left(\sqrt{\frac{\Delta_z \Gamma M}{\pi}}\frac{1}{z-M^2-i\Gamma M} \right)+\log(f(z))\Bigg] \nonumber \\
	=& -f(z)\log\left(\sqrt{\frac{\Delta_z \Gamma M}{\pi}}\frac{1}{2i\Gamma M} \right)\Big|_{z=M^2+i\Gamma M}-f(z)\log\left(\sqrt{\frac{\Delta_z \Gamma M}{\pi}}\frac{1}{-2i\Gamma M} \right)\Big|_{z=M^2-i\Gamma M}\nonumber \\
	&-f(z)\log(f(z))\Big|_{z=M^2+i\Gamma M}-\int_{\bigcup_i \mathcal{C}_i}f(z)\mathcal{P}_{\Gamma,M}(z)\log(f(z)) \nonumber \\
	\approx& -f(M^2) \log\left( \frac{\Delta_z}{4\pi \Gamma M}f(M^2)\right) +\mathcal{O}\left(\frac{\Gamma} {M}\right) \, , \label{eq:Cauchy_entropy}
\end{align}
where $\mathcal{P}_{\Gamma,M}(z)$ is separated into two parts with poles in upper and lower half planes respectively, and $\mathcal{C}_i$ are appropriate contours enclosing the branch cuts associated to $z_i$, whose contributions are suppressed by an order estimation $\mathcal{P}_{\Gamma,M}(z)\propto \Gamma M$ for $z\in \bigcup_i \mathcal{C}_i$. Finally, by neglecting the suppressed contribution in $z\notin[z_L,z_R]$, we have 
\begin{align}
	-\int^{z_R}_{z_L}dz\, f(z) \mathcal{P}_{\Gamma,M}(z) \log\left(\Delta_z f(z) \mathcal{P}_{\Gamma,M}(z)\right) \approx& -\int^{+\infty}_{-\infty}dz\, f(z) \mathcal{P}_{\Gamma,M}(z) \log\left(\Delta_z f(z) \mathcal{P}_{\Gamma,M}(z)\right) \nonumber \\
	&+\mathcal{O}\left(\frac{\Gamma}{M}\right) \, .
\end{align}

It is noteworthy that there is a factor of $4$ difference in the logarithm of (\ref{eq:Cauchy_entropy}) if we naively approximate the integral as the value at the sharp peak $z=M^2$, as used in \cite{Lello:2013bva} for calculating the entanglement entropy by decay:
\begin{align}
	&-\int^{+\infty}_{-\infty}dz\, f(z) \mathcal{P}_{\Gamma,M}(z) \log\left(\Delta_z f(z) \mathcal{P}_{\Gamma,M}(z)\right)  \nonumber \\
	\stackrel{?}{\approx}& -\int^{+\infty}_{-\infty}dz\, f(z) \delta(z-M^2) \log\left(\Delta_z f(z) \mathcal{P}_{\Gamma,M}(z)\right)+\mathcal{O}\left(\frac{\Gamma}{M}\right) \nonumber \\
	=&-f(M^2) \log\left( \frac{\Delta_z}{\pi \Gamma M}f(M^2)\right)\, . \label{eq:wrong_Cauchy_entropy}
\end{align}
The reason why the result of (\ref{eq:Cauchy_expectation}) does not apply for $f(z)\to -f(z)  \log\left(\Delta_z f(z) \mathcal{P}_{\Gamma,M}(z)\right)$ is that the poles of $\mathcal{P}_{\Gamma,M}(z)$ also produce branch cuts by $\log\left(\Delta_z f(z) \mathcal{P}_{\Gamma,M}(z)\right)$, so the direct deformation of integral contour does not work. The resolution is to separate $\mathcal{P}_{\Gamma,M}(z)$ into the two parts with poles only in the upper and lower-half planes respectively, as shown in (\ref{eq:Cauchy_entropy}), such that the singularities in the logarithm can be avoided. In Appendix~\ref{sec:factor_4}, we also numerically compute the entanglement entropy of decay in \cite{Lello:2013bva}, showing that the formula (\ref{eq:Cauchy_entropy}) is correct.

\subsection{The $n$-body integrals and reduced density matrix by the phase-space recursive relation} \label{sec:n-particle-recursion}
\begin{figure}
	\centering
	\includegraphics[width=0.6\textwidth]{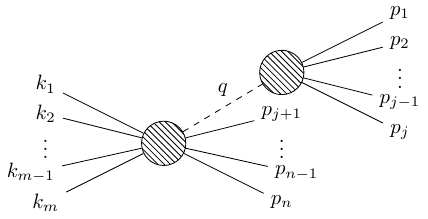}
	\caption{Applying the decomposition of phase-space integral (\ref{eq:phase_space_decompose}) to the diagram of (\ref{eq:M_decompose}), where the left and right blobs corresponding to $i\mathcal{M}(m\to n-j+1)$ and $i\mathcal{M}(1\to j)$ respectively. \label{fig:apply_decompose}}
\end{figure}

For the case of multiple phase-space integrals, the part related to the resonance structure can be decomposed as follows by the recursive relation \cite{ParticleDataGroup:2024cfk,Jing:2020tth}
\begin{align}
	\int d\Pi_n(K;p_1,\dots,p_n)&=\int \frac{dq^2}{2\pi} d\Pi_{n-j+1}(K;q,p_{j+1},\dots,p_n) d\Pi_j(q;p_1,\dots,p_j) \, , \label{eq:phase_space_decompose}
\end{align}
where $q$ is a four-vector such that the integration variable $q^2$, acting like an invariant mass, determines its zero component $q^0=\sqrt{q^2+\bar{q}^2}$. This decomposition is particularly useful if the amplitude can be decomposed as
\begin{align}
	\left|\mathcal{M}(m\to n)\right|^2&=\frac{1}{(q^2-M^2)^2+\Gamma^2M^2}\left|\mathcal{M}(1\to j)\right|^2 \left|\mathcal{M}(m\to n-j+1)\right|^2 \nonumber \\
	&=\frac{\pi}{\Gamma M} \mathcal{P}_{\Gamma,M}(q^2) \left|\mathcal{M}(1\to j)\right|^2 \left|\mathcal{M}(m\to n-j+1)\right|^2\, , \label{eq:M_decompose}
\end{align}
where the first $j$ particles come from the decay of a heavy particle, so the $j$-body phase-space integral in (\ref{eq:phase_space_decompose}) fixes $\sum_{i=1}^j p_i=q$, as shown in Figure~\ref{fig:apply_decompose}.

With (\ref{eq:phase_space_decompose}) and (\ref{eq:M_decompose}), the resonant part in the $n$-body phase-space integral is reduced to a one-dimensional integral, so that we can apply the approximation (\ref{eq:Cauchy_expectation}) with the Cauchy distribution in $q^2$:
\begin{align}
	&\mathcal{I}^{(m \to n)} \nonumber \\
	=&\int d\Pi_n \left|\mathcal{M}(m\to n)\right|^2 \nonumber \\
	=&\int \frac{dq^2}{2\pi} \frac{\pi}{\Gamma M} \mathcal{P}_{\Gamma,M}(q^2) \int d\Pi'_{n-j+1}(K';q',p'_{j+1},\dots,p'_n)\left|\mathcal{M}(m\to n-j+1)\right|^2 \nonumber \\
	&\times \int d\Pi'_j (q';p'_1,\dots,p'_j)\left|\mathcal{M}(1\to j)\right|^2  \nonumber \\
	\approx& \frac{\int d\Pi'_j (q';p'_1,\dots,p'_j)\left|\mathcal{M}(1\to j)\right|^2}{2\Gamma M} \int d\Pi'_{n-j+1}(K';q',p'_{j+1},\dots,p'_n)\left|\mathcal{M}(m\to n-j+1)\right|^2   \Big|_{q'^2=M^2} \nonumber \\
	&+\mathcal{O}\left(\frac{\Gamma}{M}\right) \nonumber \\
	=&\frac{\Gamma_{1\to j}(M^2)}{\Gamma}\mathcal{I}^{(m\to n-j+1)}(M^2) \, , \label{eq:Imn_Imn-j+1_recursion}
\end{align}
where the frame is boosted to the one with $\vec{q'}=\vec{0}$, so the integrand depends only on the one-dimensional Lorentz-invariant variable $q^2$, and in the last line the two quantities are the decay rate of $1\to j$ process 
\begin{align}
	\Gamma_{1\to j}(q^2)&=\frac{\int d\Pi'_j (q';p'_1,\dots,p'_j)\left|\mathcal{M}(1\to j)\right|^2}{2 M} \nonumber \\
    &=\frac{\int d\Pi_j(q;p_1,\dots,p_j)\left|\mathcal{M}(1\to j)\right|^2}{2M} \nonumber \\
    &=\frac{\mathcal{I}^{(1\to j)}(q^2)}{2M}
    \, ,
\end{align}
and the phase-space integral of the $m\to n-j+1$ scattering
\begin{align}
	\mathcal{I}^{(m\to n-j+1)}(q^2)&=\int d\Pi'_{n-j+1}(K';q',p'_{j+1},\dots,p'_n)\left|\mathcal{M}(m\to n-j+1)\right|^2   \nonumber \\
	&=\int d\Pi_{n-j+1}(K;q,p_{j+1},\dots,p_n)\left|\mathcal{M}(m\to n-j+1)\right|^2   \, , \label{eq:Im_n-j1_q2}
\end{align}
respectively. It is noteworthy that $\mathcal{I}^{(m\to n)}$ and $\mathcal{I}^{(m\to n-j+1)}$ are related by a branching ratio $\Gamma_{1\to j}(M^2)/\Gamma$ in (\ref{eq:Imn_Imn-j+1_recursion}), and the integrand vanishes for $q^2>K^2=E_t^2$, justifying the requirement at $q^2\to\infty$ for applying the approximation (\ref{eq:Cauchy_expectation}).

Now we consider the reduced density matrix of decay products, in a general inelastic scattering from $m$ initial particles to $n$ final particles.\footnote{There is no forward limit in this case, unlike the elastic scattering.}  Similar to the previous $2\to 2$ scattering, the final out state is defined through a projection to the $n$-particle state:
\begin{align}
	|f^{(m\to n)}\rangle 
	&= \frac{1}{\sqrt{\mathcal{N}^{(m \to n)} }}\int d\Pi_n(K;p_1,\dots,p_n)\, i \mathcal{M}\left(m \to n\right) |p_{1}p_{2}\dots p_n\rangle \, ,
\end{align}
and the normalization condition $\langle f^{(m\to n)}|f^{(m \to n)}\rangle =1$ implies
\begin{align}
	\mathcal{N}^{(m \to n)} 
	&=\int d\Pi_n(K;p_1,\dots,p_n)\, (2\pi)^4\delta^4\left(K - \sum_{i=1}^n p_i\right)  \left|\mathcal{M}(m\to n)\right|^2 \nonumber \\
    &=\mathcal{I}^{(m\to n)}(2\pi)^4\delta^4(0)\, .
\end{align}

For convenience, we shorten the label of quantities of the decay products, the particles from $1$ to $j$, as $Q^{(m\to n)}_{12\dots j}$=$Q^{(m\to n)}_{1-j}$. The reduced density matrix of the decay products obtained by tracing out the particles from $j+1$ to $n$, $\rho_{1-j}^{(m\to n)}=\rho_{12\dots j}^{(m\to n)}$, can be calculated easily with the phase-space recursive relation (\ref{eq:phase_space_decompose}):
\begin{align}
	&\rho_{1-j}^{(m \to n)} \nonumber \\
	=&\int \frac{d^3p_{j+1}}{(2\pi)^3 2E_{j+1}}\dots \frac{d^3p_{n}}{(2\pi)^3 2E_{n}} \langle p_{j+1}\dots p_n|f^{(m\to n)}\rangle\langle f^{(m\to n)} |p_{j+1}\dots p_n \rangle \nonumber \\
	=& \int d\Pi_n(K;p_1,\dots,p_n) \int\prod_{i=1}^j\frac{d^3p''_i}{(2\pi)^3 2E_i''}
	\nonumber \\
	&\times \mathcal{M}(m\to n){\mathcal{M}''}^*(m\to n)  (2\pi)^4\delta^4\left(K-\sum_{i=1}^n  p_i''\right) \frac{|p_1 \dots p_j\rangle \langle p_1'' \dots p_j''|}{\mathcal{N}^{(m\to n)}} \nonumber \\
	=&\frac{1}{\mathcal{N}^{(m\to n)}}\int \frac{dq^2}{2\pi}\frac{\pi}{\Gamma M}\mathcal{P}_{\Gamma,M}(q^2) \int d\Pi_{n-j+1}(K;q,p_{j+1},\dots,p_n)\left|\mathcal{M}(m\to n-j+1)\right|^2 \nonumber \\
	&\times \left(\int d\Pi_j(q;p_1,\dots,p_j)\mathcal{M}(1\to j)|p_1,\dots,p_j\rangle\right)\left(\int d\Pi''_j(q;p''_1,\dots,p''_j){\mathcal{M}''}^*(1\to j)\langle p''_1,\dots,p''_j|\right) \nonumber \\
	=&\int \frac{dq^2}{2\pi}\frac{\pi}{\Gamma M}\mathcal{P}_{\Gamma,M}(q^2) \frac{\mathcal{I}^{(1\to j)}(q^2)}{\mathcal{I}^{(m \to n)}} \Theta\left(q^2-\left(\sum_{i=1}^j m_i\right)^2\right) \nonumber \\
	&\times\int d\Pi_{n-j+1}(K;q,p_{j+1},\dots,p_n)\left|\mathcal{M}(m\to n-j+1)\right|^2 |\tilde{\psi}^{(m\to n)}_{1-j}(q)\rangle \langle \tilde{\psi}^{(m\to n)}_{1-j}(q)| \, , \label{eq:rho_1j}
\end{align}
where $\mathcal{M}''(m\to n)$ means depending on the momenta $p_1'',\dots,p_n''$, and the orthonormal $j$-particle basis is
\begin{align}
	&|\tilde{\psi}^{(m\to n)}_{1-j}(q)\rangle \nonumber \\
    =&\sqrt{\frac{1}{\int d\Pi_j \left|\mathcal{M}(1\to j)\right|^2(2\pi)^4\delta^4\left(q-\sum_{i=1}^jp_i\right)}}\left(\int d\Pi_j(q;p_1,\dots,p_j)\mathcal{M}(1\to j)|p_1,\dots,p_j\rangle\right) \nonumber \\
    =&\sqrt{\frac{1}{\mathcal{I}^{(1\to j)}(q^2)(2\pi)^4\delta^4\left(0\right)}}\left(\int d\Pi_j(q;p_1,\dots,p_j)\mathcal{M}(1\to j)|p_1,\dots,p_j\rangle\right) \, ,
\end{align}
with vanishing inner product if either ${\vec{q}}$ or $q^2$ is different, and its existence requires the step function of $q^2\geq \left(\sum_{i=1}^j m_i\right)^2$. It is noteworthy that the reduced density matrix (\ref{eq:rho_1j}) can be constructed by suitably marginalizing the phase-space distribution of final particles to $(q^2,\vec{q})$, as shown in the following with (\ref{eq:phase_space_decompose}) and (\ref{eq:M_decompose}):
\begin{align}
    &\int d\Pi_n(K;p_1,\dots,p_n)\frac{\left|\mathcal{M}(m\to n)\right|^2}{\mathcal{I}^{(m\to n)}} \nonumber\\
    =&\int \frac{dq^2}{2\pi} d\Pi_{n-j+1}(K;q,p_{j+1},\dots,p_n) d\Pi_j(q;p_1,\dots,p_j) \frac{\left|\mathcal{M}(m\to n)\right|^2}{\mathcal{I}^{(m\to n)}} \nonumber \\
    =&\int \frac{dq^2}{2\pi} d\Pi_{n-j+1}(K;q,p_{j+1},\dots,p_n) d\Pi_j(q;p_1,\dots,p_j) \nonumber \\
    &\times \frac{\pi}{\Gamma M} \mathcal{P}_{\Gamma,M}(q^2) \left|\mathcal{M}(1\to j)\right|^2 \frac{\left|\mathcal{M}(m\to n-j+1)\right|^2}{\mathcal{I}^{(m\to n)}}\Theta\left(q^2-\left(\sum_{i=1}^jm_i\right)^2\right) \nonumber \\
    =&\int \frac{dq^2}{2\pi}\frac{\pi}{\Gamma M}\mathcal{P}_{\Gamma,M}(q^2) \frac{\mathcal{I}^{(1\to j)}(q^2)}{\mathcal{I}^{(m \to n)}} \Theta\left(q^2-\left(\sum_{i=1}^j m_i\right)^2\right) \nonumber \\
    &\times \int d\Pi_{n-j+1}(K;q,p_{j+1},\dots,p_n)\left|\mathcal{M}(m\to n-j+1)\right|^2 
    \, ,
\end{align}
where the step function constrained by kinematics is explicitly written down in the second equality, giving a marginal probability distribution for each $(q^2,\vec{q})$. Therefore, the entanglement entropy and its features discussed in the following Section~\ref{sec:structure_decay_onshell}, based on the reduced density matrix (\ref{eq:rho_1j}), can in principle be probed by analyzing the particle detector's data.

\subsection{The entanglement structure by decay and the on-shell approximation}\label{sec:structure_decay_onshell}
The reduced density matrix $\rho^{(m \to n)}_{1-j}$ (\ref{eq:rho_1j}) is a mixture of quantum states, written as a linear combination over different $q^2$, and this is easily seen as follows:
\begin{align}
	\rho^{(m \to n)}_{1-j}=\int dq^2 P^{(m \to n)}_{1-j}(q^2) \sigma^{(m \to n)}_{1-j}(q^2) \, , \label{eq:rho_1j_mixture}
\end{align}
where the probability distribution of $q^2$ is\footnote{$\mathcal{I}^{(m\to n-j+1)}(q^2)$ is a Lorentz-invariant function of $q^2$ which can also be shown as follows. We can first integrate $p_{j+1},\dots,p_n$ with $\delta^4(K-q-\sum_{i=j+1}^np_i)$, and the result should depend on the Lorentz scalar $(K-q)^2=E_t^2+q^2-2E_t\sqrt{q^2+\bar{q}^2}$. Then we finish the last integral of $\int d^3q$, and the result is thus a function depending on $q^2$. We will see this fact in (\ref{eq:3_body_q_3_4}) and (\ref{eq:I24_integral}), where a concrete model of  $2\to 4$ scattering is studied in Section~\ref{sec:2-4_example}.}
\begin{align}
	P^{(m \to n)}_{1-j}(q^2)= \mathcal{P}_{\Gamma,M}(q^2)\frac{\mathcal{I}^{(1\to j)}(q^2) \mathcal{I}^{(m \to n-j+1)}(q^2)}{2\Gamma M\mathcal{I}^{(m \to n)}}\Theta\left(q^2-\left(\sum_{i=1}^j m_i\right)^2\right) \, , \label{eq:P_1j_mixture}
\end{align}
with $\int dq^2 P^{(m \to n)}_{1-j}(q^2)=1$, and the reduced density matrix of $q^2$ is
\begin{align}
	 &\sigma^{(m \to n)}_{1-j}(q^2) \nonumber \\
	 =&\int d\Pi_{n-j+1}(K;q,p_{j+1},\dots,p_n)\frac{\left|\mathcal{M}(m\to n-j+1)\right|^2}{\mathcal{I}^{(m \to n-j+1)}(q^2)}  |\tilde{\psi}^{(m\to n)}_{1-j}(q)\rangle \langle \tilde{\psi}^{(m\to n)}_{1-j}(q)| \nonumber \\
	 =&\sum_{\vec{q}}\frac{1}{2q^0 V} \int \prod_{i=j+1}^n\frac{d^3p_i}{(2\pi)^32E_i}(2\pi)^4\delta^4\left(K-q-\sum_{i=j+1}^n p_i\right)\frac{\left|\mathcal{M}(m\to n-j+1)\right|^2}{\mathcal{I}^{(m \to n-j+1)}(q^2)}  \nonumber \\
	 &\times |\tilde{\psi}^{(m\to n)}_{1-j}(q)\rangle \langle \tilde{\psi}^{(m\to n)}_{1-j}(q)| \nonumber \\
	 =&\sum_{\vec{q}}\frac{1}{2q^0 V} f^{(m\to n)}_{1-j}(q)|\tilde{\psi}^{(m\to n)}_{1-j}(q)\rangle \langle \tilde{\psi}^{(m\to n)}_{1-j}(q)| \, . \label{eq:sigma_1j_mixture}
\end{align}

To calculate its entropy, we need the discretization factor $\Delta_{q^2}$ for the continuous variable $q^2$, determined by the relation
\begin{align}
	\int\frac{dq^2}{2\pi}\int\frac{d^3q}{(2\pi)^3 2q^0} f(q^0,\vec{q}) &= \int \frac{d^4 q}{(2\pi)^4} \int dq^2 \delta((q^0)^2-\bar{q}^2-q^2) \Theta(q^0) f(q^0,\vec{q}) \nonumber \\
	&=\int \frac{d^4 q}{(2\pi)^4}  f(q^0,\vec{q}) \, ,
\end{align}
for an arbitrary function $f(q^0,\vec{q})$ non-vanishing only for $q^0\geq 0$, and converting these to discretized sums gives
\begin{align}
	\sum_{q^2,\vec{q}}\frac{\Delta_{q^2}}{2\pi}\frac{1}{2q^0V}f(q^0,\vec{q})=\sum_{q^0,\vec{q}}\frac{1}{VT}f(q^0,\vec{q}) \, .
\end{align}
Since each point $(q^0,\vec{q})$ is related to $(\sqrt{q^2+\bar{q}^2},\vec{q})$, their measures are equal, implying that 
\begin{align}
	\frac{\Delta_{q^2}}{2\pi}\frac{1}{2q^0}=\frac{1}{T} \, . \label{eq:deltaq2_discretization}
\end{align}
With (\ref{eq:rho_1j_mixture}), (\ref{eq:sigma_1j_mixture}) and (\ref{eq:deltaq2_discretization}), the entanglement entropy is
\begin{align}
	S_{EE}\left(\rho^{(m \to n)}_{1-j}\right)&=-\int dq^2 \frac{d^3q}{(2\pi)^3 2 q^0}P^{(m \to n)}_{1-j}(q^2) f^{(m\to n)}_{1-j}(q) \log\left(\frac{2\pi}{VT} P^{(m \to n)}_{1-j}(q^2) f^{(m\to n)}_{1-j}(q) \right) \nonumber \\
	&=h\left(P^{(m \to n)}_{1-j}(q^2)\right) +\int dq^2 P^{(m \to n)}_{1-j}(q^2) s\left(f^{(m\to n)}_{1-j}(q) \right)+\log\left(\frac{VT}{2\pi}\right) \, , \label{eq:SEE_1j_mixture}
\end{align}
where the normalizations $\int dq^2P^{(m \to n)}_{1-j}(q^2)=\int\frac{d^3q}{(2\pi)^3 2q^0 }f^{(m\to n)}_{1-j}(q)=1$ are applied, and $h$ and $s$ are differential (continuous) entropies
\begin{align}
	h\left(P^{(m \to n)}_{1-j}(q^2)\right) &=-\int dq^2 P^{(m \to n)}_{1-j}(q^2) \log\left(P^{(m \to n)}_{1-j}(q^2) \right) \, , \label{eq:h_1j_mixture} \\
	s\left(f^{(m\to n)}_{1-j}(q) \right) &= - \int \frac{d^3q}{(2\pi)^3 2q^0} f^{(m\to n)}_{1-j}(q) \log\left(f^{(m\to n)}_{1-j}(q)\right) \label{eq:s_1j_state}\, .
\end{align}
It is noteworthy that (\ref{eq:SEE_1j_mixture}) is analogic to (\ref{eq:S_mixture}) in $2\to 2$ scattering, and here the mixture part (\ref{eq:h_1j_mixture}) is attributed to the continuous probability distribution $P^{(m\to n)}_{1-j}(q^2)$ involving the heavy-field propagator, whereas the microscopic state with density matrix $\sigma^{(m \to n)}_{1-j}(q^2)$ contributes to (\ref{eq:s_1j_state}).

In the case with $\Gamma \ll M$ and high enough total energy $E_t\gtrsim M$, the probability distribution $P^{(m\to n)}_{1-j}(q^2)$ (\ref{eq:P_1j_mixture}) has a sharp peak at $q^2=M^2$, and the on-shell approximations by the Cauchy distribution (\ref{eq:Cauchy_expectation}) and (\ref{eq:Cauchy_entropy}) are applicable:
\begin{align}
	&h\left(P^{(m \to n)}_{1-j}(q^2)\right) \nonumber \\
	\approx& -\frac{\mathcal{I}^{(1\to j)}(M^2) \mathcal{I}^{(m \to n-j+1)}(M^2)}{2\Gamma M\mathcal{I}^{(m \to n)}} \log\left(\frac{1}{4\pi\Gamma M}\frac{\mathcal{I}^{(1\to j)}(M^2) \mathcal{I}^{(m \to n-j+1)}(M^2)}{2\Gamma M\mathcal{I}^{(m \to n)}}\right) +\mathcal{O}\left(\frac{\Gamma}{M}\right) \nonumber \\
	\approx& \log\left(4\pi\Gamma M\right) +\mathcal{O}\left(\frac{\Gamma}{M}\right)  \, . \label{eq:h_1j_on-shell}
\end{align}
where (\ref{eq:Imn_Imn-j+1_recursion}) is applied to relate $\mathcal{I}^{(m \to n-j+1)}(M^2)$ and $\mathcal{I}^{(m \to n)}$, and 
\begin{align}
	\int dq^2 P^{(m \to n)}_{1-j}(q^2) s\left(f^{(m\to n)}_{1-j}(q) \right) \approx s\left(f^{(m\to n)}_{1-j}(q) \right)\Big|_{q^2=M^2} +\mathcal{O}\left(\frac{\Gamma}{M}\right) \, .
\end{align}
Therefore, the entanglement entropy (\ref{eq:SEE_1j_mixture}) has the following on-shell approximation
\begin{align}
	S_{EE}\left(\rho^{(m \to n)}_{1-j}\right) \approx \log\left(4\pi\Gamma M\right)+s\left(f^{(m\to n)}_{1-j}(q) \right)\Big|_{q^2=M^2}+\log\left(\frac{VT}{2\pi}\right) +\mathcal{O}\left(\frac{\Gamma}{M}\right) \, , \label{eq:SEE_1j_on-shell}
\end{align}
where the decay rate is evaluated at resonance $\Gamma=\Gamma(M^2)$, as the on-shell approximation relies only on $\mathcal{P}_{\Gamma,M}(z)$ around its poles.

From (\ref{eq:SEE_1j_on-shell}), the entanglement entropy, dominated by an on-shell heavy particle, is universally suppressed by small $\Gamma$, and this originates from the mixture part governed by the heavy-field propagator, as demonstrated by (\ref{eq:h_1j_on-shell}). The second part with $s\left(f^{(m\to n)}_{1-j}(q) \right)$ is independent to $\Gamma$, whereas it is related to the number of state allowed by kinematics, as expected for inelastic scattering. At the low-energy regime up to $E_t\lesssim M$, the theory and the corresponding $S_{EE}\left(\rho^{(m \to n)}_{1-j}\right)$ are well approximated by an EFT independent to $\Gamma$, and thus the entanglement entropy is expected to have a dip for matching the suppressed on-shell approximation at $E_t\gtrsim M$. We will demonstrate this feature in a concrete $2\to 4$ scattering model in Section~\ref{sec:example}.

On the other hand, the third term in (\ref{eq:SEE_1j_on-shell}) indicates that the entanglement entropy increases by the spacetime volume $VT$, provided that $s\left(f^{(m\to n)}_{1-j}(q) \right)$ does not include any such a factor to cancel it. It is general true except for the case with $j=n-1$, where $f^{(m \to n)}_{1-(n-1)}$ is proportional to the interaction time $T$. As a result, the entanglement entropy scales with $\log(V)$ only. The following Section~\ref{sec:jn-1} discusses such a special case.

\subsection{Entanglement entropy with $j=n-1$} \label{sec:jn-1}
\begin{figure}
	\centering
	\includegraphics[width=\textwidth]{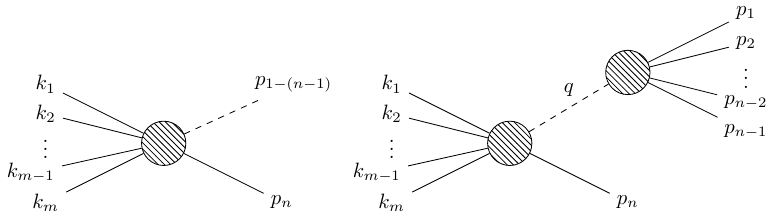}
	\caption{The case with $j=n-1$. {\bf Left:} A general inelastic $m \to 2$ scattering, represents by the part $i\mathcal{M}(m\to2)$, where a heavy particle with momentum $p_{1-(n-1)}$ is produced. {\bf Right:} The heavy particle further decays into $n-1$ light particles, forming a subsystem described by $\rho^{(m \to n)}_{1-(n-1)}$, and the whole scattering process is represented by $i\mathcal{M}(m \to n)$.\label{fig:j_n-1}}
\end{figure}

For the case with $j=n-1$ in (\ref{eq:rho_1j}), corresponding to Figure~\ref{fig:j_n-1}, the reduced density matrix is
\begin{align}
	\rho_{1-(n-1)}^{(m \to n)} 
	=&\int \frac{dq^2}{2\pi}\frac{\pi}{\Gamma M}\mathcal{P}_{\Gamma,M}(q^2) \int \frac{d^3q}{(2\pi)^32q^0}\frac{d^3p_n}{(2\pi)^32E_n}(2\pi)^4\delta^4(K-q-p_n)\left|\mathcal{M}(m\to 2)\right|^2  \nonumber \\
	&\times\frac{\mathcal{I}^{(1\to n-1)}(q^2)}{\mathcal{I}^{(m \to n)}}\Theta\left(q^2-\left(\sum_{i=1}^{n-1} m_i\right)^2\right) |\tilde{\psi}^{(m\to n)}_{1-(n-1)}(q)\rangle \langle \tilde{\psi}^{(m\to n)}_{1-(n-1)}(q)| \nonumber \\
	=& \int \frac{d^3q}{(2\pi)^3} \int \frac{dq^2}{2\pi}\frac{\pi}{\Gamma M}\mathcal{P}_{\Gamma,M}(q^2)(2\pi)\frac{\delta\left(q^2-z(\bar{q})\right)}{2E_n}\left|\mathcal{M}(m\to 2)\right|^2  \nonumber \\
	&\times\frac{\mathcal{I}^{(1\to n-1)}(q^2)}{\mathcal{I}^{(m \to n)}}\Theta\left(q^2-\left(\sum_{i=1}^{n-1} m_i\right)^2\right) |\tilde{\psi}^{(m\to n)}_{1-(n-1)}(q)\rangle \langle \tilde{\psi}^{(m\to n)}_{1-(n-1)}(q)| \nonumber \\
	=&\int \frac{d^3q}{(2\pi)^32E_{n}} \mathcal{F}^{(m \to n)}_{1-(n-1)}(q) |\tilde{\psi}^{(m\to n)}_{1-(n-1)}(q)\rangle \langle \tilde{\psi}^{(m\to n)}_{1-(n-1)}(q)| \, ,  \label{eq:rho_mn_1-n-1}
\end{align}
where the delta function in the second equality comes from
\begin{align}
	\frac{\delta(E_t-q^0-E_n)}{2q^0} &= \delta(q^2-z(\bar{q})) \, , \label{eq:deltaE_to_deltaq2}
\end{align}
with
\begin{align}
	z(\bar{q})&=E_t^2+m_n^2-2E_t \sqrt{m_n^2+\bar{q}^2} \, , \label{eq:z_qbar_j_n-1}
\end{align}
for fixed $\vec{q}$, and 
\begin{align}
	\mathcal{F}^{(m \to n)}_{1-(n-1)}(q)=\frac{\pi}{\Gamma M}\mathcal{P}_{\Gamma,M}(z)\left|\mathcal{M}(m\to 2)\right|^2\frac{\mathcal{I}^{(1\to n-1)}(z)}{\mathcal{I}^{(m \to n)}}\Theta\left(z-\left(\sum_{i=1}^{n-1} m_i\right)^2\right) \, . \label{eq:Fmn_1_n-1}
\end{align}
Another way to understand (\ref{eq:deltaE_to_deltaq2}) is that the integrand of the reduced density matrix (\ref{eq:rho_mn_1-n-1}) is proportional to $T=2\pi\delta(E_t-q^0-E_n)$, canceling with $\Delta_{q^2}=(\delta(q^2-z(\bar{q})))^{-1}=4\pi q^0/T$. This will lead to the entanglement entropy independent of $T$.

The corresponding entanglement entropy can be approximated with the Cauchy distribution (\ref{eq:Cauchy_entropy})
\begin{align}
	S_{EE}\left(\rho_{1-(n-1)}^{(m \to n)} \right)
	=&-\int \frac{d^3q}{(2\pi)^32E_{n}} \mathcal{F}^{(m \to n)}_{1-(n-1)}(q) \log\left(\frac{\mathcal{F}^{(m \to n)}_{1-(n-1)}(q)}{2E_{n}V}\right) \nonumber \\
	=&-\int d\Omega \left[\int \frac{dz}{(2\pi)^3} \frac{\bar{q}^2(z)}{2E_n(z)}\frac{d\bar{q}}{dz}\mathcal{F}^{(m \to n)}_{1-(n-1)}(q) \log\left(\frac{\mathcal{F}^{(m \to n)}_{1-(n-1)}(q)}{2E_{n}(z)V}\right) \right] \nonumber \\
	\approx& -\int d\Omega \Bigg[ \frac{1}{(2\pi)^2} \frac{\bar{p}_{CM}(E_t;\sqrt{z},m_n)}{4E_t} \frac{ \Gamma_{1\to n-1}}{\Gamma} \frac{\left|\mathcal{M}(m\to 2)\right|^2}{\mathcal{I}^{(m \to n)}} \nonumber \\
	&\times \log\left(\frac{1}{4\Gamma ME_{n}(z)V}\frac{ \Gamma_{1\to n-1}}{\Gamma} \frac{\left|\mathcal{M}(m\to 2)\right|^2}{\mathcal{I}^{(m \to n)}}\right) \Bigg] \Bigg|_{z=M^2} +\mathcal{O}\left(\frac{\Gamma}{M}\right) \, , \label{eq:SEE_1n-1_onshell_angle}
\end{align}
where the solid angle $\Omega$ indicates the direction of $\vec{q}$, and we use the facts that $\mathcal{I}^{(1 \to n-1)}(M^2)=2M\Gamma_{1\to n-1}(M^2)$ as well as
\begin{align}
    \frac{\bar{q}^2(z)}{2E_n(z)}\frac{d\bar{q}}{dz}=-\frac{\bar{p}_{CM}(E_t;\sqrt{z},m_n)}{4E_t} \, ,
\end{align}
where
\begin{align}
    E_n(z)&=\frac{E_t^2+m_n^2-z}{2E_t}\, . \label{eq:z_En_def} 
\end{align}
In the special case that $\mathcal{M}(m \to 2)$ independent to the angle, the result can be further simplified as
\begin{align}
	S_{EE}\left(\rho_{1-(n-1)}^{(m \to n)} \right) \approx \log\left(\Gamma \frac{\bar{p}_{CM}(E_t;M,m_n)ME_nV}{\pi E_t}\right)  \, , \label{eq:SEE_1n-1_onshell}
\end{align}
where the we apply the normalization in the on-shell limit
\begin{align}
    {\rm Tr}\rho_{1-(n-1)}^{(m \to n)}& \approx \int d\Omega  \frac{1}{(2\pi)^2} \frac{\bar{q}^2(z)}{2E_n(z)}\frac{d\bar{q}}{dz} \frac{ \Gamma_{1\to n-1}}{\Gamma} \frac{\left|\mathcal{M}(m\to 2)\right|^2}{\mathcal{I}^{(m \to n)}}\Big|_{z=M^2}\nonumber \\
    &=\frac{\bar{p}_{CM}(E_t;M,m_n)}{4\pi E_t}\frac{ \Gamma_{1\to n-1}}{\Gamma} \frac{\left|\mathcal{M}(m\to 2)\right|^2}{\mathcal{I}^{(m \to n)}} \nonumber\\
    &=1 \, .
\end{align}
In the ultra-relativistic limit, (\ref{eq:SEE_1n-1_onshell}) implies the tendency of the entanglement entropy \begin{align}
S_{EE}\left(\rho_{1-(n-1)}^{(m \to n)} \right)\sim \log\left(\Gamma \frac{E_t M V}{4\pi}\right) \, ,
\end{align}
which scales as $\log(\Gamma E_t)$. Since the on-shell contribution is also suppressed by small $\Gamma$, a sharp dip of entanglement entropy at $E_t\gtrsim M$ is expected.  We will verify the feature numerically in a $2\to3$ model in Section~\ref{sec:example}, in which we also observe that the entropy of lower-energy EFT scales faster than $\log(E_t)$.

On the other hand, as a consistency check, we also calculate the entanglement entropy with the reduced density matrix of the $n$-th particle $\rho^{(m \to n)}_n$, which has to be $S_{EE}\left(\rho_{1-(n-1)}^{(m \to n)}\right)$ as the full system is in a pure state. The reduced density matrix obtained by tracing out the first $n-1$ particles is
\begin{align}
	&\rho^{(m \to n)}_n \nonumber \\
        =& \int\prod_{i=1}^{n-1}\frac{d^3p_i}{(2\pi)^32E_i} \langle p_1,\dots,p_{n-1}|f^{(m\to n)}\rangle\langle f^{(m\to n)}|p_1,\dots,p_{n-1}\rangle\nonumber \\
        =& \frac{1}{\mathcal{I}^{(m\to n)}} \int d\Pi_n(K;p_1,\dots,p_n) \left|\mathcal{M}(m\to n)\right|^2 |\tilde{p}_n\rangle \langle \tilde{p}_n|\nonumber \\
	=&\frac{1}{\mathcal{I}^{(m\to n)}}\int \frac{dq^2}{2\pi} \frac{\pi}{\Gamma M} \mathcal{P}_{\Gamma,M}(q^2) \int d\Pi_{2}(K;q,p_{n})\left|\mathcal{M}(m\to 2)\right|^2 \nonumber \\
	&\times \int d\Pi_{n-1} (q;p_1,\dots,p_{n-1})\left|\mathcal{M}(1\to n-1)\right|^2 |\tilde{p}_n\rangle \langle \tilde{p}_n| \nonumber \\
	=&\int \frac{d^3p_n}{(2\pi)^32E_n}\int \frac{dq^2}{2\pi} \frac{\pi}{\Gamma M} \mathcal{P}_{\Gamma,M}(q^2) (2\pi)\delta(q^2-z(\bar{p}_n))\left|\mathcal{M}(m\to 2)\right|^2    \frac{\mathcal{I}^{(1 \to n-1)}(q^2)}{\mathcal{I}^{(m\to n)}}|\tilde{p}_n\rangle \langle \tilde{p}_n| \nonumber \\
	=& \int \frac{d^3p_n}{(2\pi)^32E_n} \mathcal{F}_n^{(m \to n)}(p_n)|\tilde{p}_n\rangle \langle \tilde{p}_n| \, , \label{eq:rho_n_mton}
\end{align}
where $\vec{p}_n=-\vec{q}$ and $z(\bar{p}_n)$ defined similarly as (\ref{eq:z_qbar_j_n-1}) in the CM frame, so the coefficient of reduced density matrix
\begin{align}
	\mathcal{F}_n^{(m \to n)}(p_n)=\frac{\pi}{\Gamma M}\mathcal{P}_{\Gamma,M}(z) \left|\mathcal{M}(m\to 2)\right|^2   \frac{\mathcal{I}^{(1 \to n-1)}(z)}{\mathcal{I}^{(m \to n)}} \Theta\left(z-\left(\sum_{i=1}^{n-1} m_i\right)\right) \, ,
\end{align}
which is exactly equal to $\mathcal{F}_{1-(n-1)}^{(m \to n)}(q)$ (\ref{eq:Fmn_1_n-1}). Therefore, we must have 
\begin{align}
	S_{EE}\left(\rho_{1-(n-1)}^{(m \to n)} \right)=S_{EE}\left(\rho_{n}^{(m \to n)} \right) \, ,
\end{align}
justifying the $(n-1)$-particle basis $|\tilde{\psi}^{(m \to n)}_{1-(n-1)}(q)\rangle$ used in (\ref{eq:rho_mn_1-n-1}). Remarkably, (\ref{eq:rho_n_mton}) indicates that the reduced density matrix $\rho_n^{(m \to n)}$ and the corresponding entanglement entropy can be constructed by marginalizing the phase-space distribution of final particles, implying that the entanglement features may be probed experimentally.

\section{Examples: $2\to 3$ and $2\to 4$ scatterings}\label{sec:example}
In this section, we verify the on-shell entanglement features derived in Section~\ref{sec:entanglement-mton} in concrete models of $2\to 3$ and $2\to 4$ scatterings. In Section~\ref{sec:2-3model}, we first set up a simple model of $2\to 3$ scattering including an intermediate heavy particle. The entanglement entropies of its lower-energy EFT and the complete theory are numerically computed in Sections~\ref{sec:low-energy_EFT} and \ref{sec:numerical_12} respectively, and the on-shell approximation derived for the case with $j=n-1$ in Section~\ref{sec:jn-1} is verified, confirming the sharp-dip feature of entanglement entropy. In Section~\ref{sec:2-4_example}, we generalize the numerical computation to a model of $2\to 4$ scattering. The entanglement features, derived from the probabilistic mixture of final states of decay products in Section~\ref{sec:structure_decay_onshell}, are verified. 

\subsection{A model of $2\to 3$ scattering} \label{sec:2-3model}
\begin{figure}
	\centering
	\includegraphics[width=\textwidth]{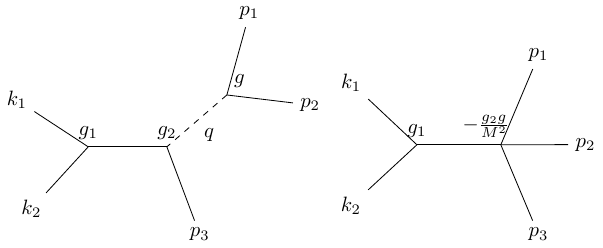}
	\caption{{\bf Left:} A simple  $2\to 3$ scattering. The solid and dashed internal lines represent particles with masses $m_\chi$ and $M$ respectively, whereas the solid external lines are light particles. {\bf Right:} An effective field theory obtained by the large $M$ expansion, and the three outgoing particles are connected by the effective vertex $-\frac{g_2g}{M^2}$. \label{fig:2-3_example}}
\end{figure}

As shown in Figure~\ref{fig:2-3_example}, the $2\to3$ scattering is described by the interaction 
\begin{align}
	\mathcal{L}_{\rm int}=g_1\phi_A\phi_B\chi+g_2\chi\sigma\phi_3+g\sigma\phi_1\phi_2 \, , \label{eq:L_int_2-3}
\end{align}
where $\phi_A$ and $\phi_B$ correspond to the initial two particles respectively, $\chi$ and $\sigma$ are intermediate particles with masses of $m_\chi$ and $M$ respectively, and finally $\phi_i$ ($i=1,2,3$) are final light particles.  The amplitude of this process is approximated by the Breit-Wigner formula
\begin{align}
	i \mathcal{M}{(2 \to 3)}&\approx i \mathcal{M}{(2 \to 2)} \frac{g}{(p_1+p_2)^2-M^2+i \Gamma M} \nonumber \\
	&=-i\frac{g_1g_2}{(k_1+k_2)^2-m_\chi^2+i\epsilon} \frac{g}{(p_1+p_2)^2-M^2+i \Gamma M} \, ,
\end{align}
where we set $\sigma$ as the heavy particle with mass $M>m_1+m_2$ for decay to happen with the rate $\Gamma$. It is noteworthy that $\mathcal{M}{(2 \to 2)}$ is a constant for a fixed initial four-momentum $K=k_1+k_2$, and therefore the non-trivial parts of the $3$-body phase-space integral and entanglement entropy are attributed to the propagator of the heavy particle. 

To be specific, we will verify numerically the on-shell approximation of entanglement entropy (\ref{eq:SEE_1n-1_onshell}):
\begin{align}
	S_{EE}\left(\rho_{12}^{(2\to 3)}\right) &\approx \log\left(\Gamma \frac{\bar{p}_{CM}(E_t;M,m_3)ME_3V}{\pi E_t}\right) \label{eq:SEE_rho_12_23_on-shell} \, ,
\end{align}
when the total energy $E_t$ is sufficient for making the intermediate heavy particle on-shell. For the low-energy behavior, we also compare with the EFT obtained by large $M$ expansion.

In Section~\ref{sec:low-energy_EFT}, we first derive the low-energy behaviors in an EFT, where we note that entanglement entropy typically grows much faster than the on-shell one (\ref{eq:SEE_rho_12_23_on-shell}). We then numerically compute $S_{EE}\left(\rho_{12}^{(2\to 3)}\right)$ in Section~\ref{sec:numerical_12}. From the numerical results, we verify the characteristic dip in entanglement entropy expected in Section~\ref{sec:structure_decay_onshell}, while transiting from the low-energy EFT to the high-energy regime well described by the on-shell approximation.

\subsection{The low-energy EFT for the $2\to 3$ scattering} \label{sec:low-energy_EFT}
For the EFT obtained by large $M$ expansion, the leading-order amplitude is 
\begin{align}
	i\mathcal{M}_{EFT}(2\to 3)\approx -i\mathcal{M}(2\to 2)\frac{g}{M^2} \, . \label{eq:EFT_amplitude}
\end{align}
We will show that the entanglement entropy of such an EFT is not suppressed by small coupling $g$ or decaying rate $\Gamma$, and its ultra-relativistic limit grows much faster than the one of the completed theory.

The calculation of the reduced density matrix $\rho_1^{EFT}$ can be greatly simplified with the constant EFT amplitude (\ref{eq:EFT_amplitude}) and by boosting to the frame with $\vec{p'}_2+\vec{p'}_3 =\vec{0}$, such that
\begin{align}
	p_2'+p_3'&=\left(\sqrt{(p_1'-K')^2},\vec{0}\right) \nonumber \\
	&=\left(\sqrt{(p_1-K)^2},\vec{0}\right)  \, . \label{eq:EFT_prime_frame}
\end{align}
With this, the coefficient $\mathcal{F}^{EFT}_1(p_1)$ of reduced density matrix is proportional to the 2-body phase-space integral in the center-of-mass frame for $p_2'$ and $p_3'$, with $\sqrt{(p_1-K)^2}$ being the total energy:
\begin{align}
	\mathcal{F}^{EFT}_1(p_1)&=\frac{1}{\mathcal{I}_{EFT}^{(2 \to 3)}} \int \frac{d^3p_2}{(2\pi)^32E_2}\frac{d^3p_3}{(2\pi)^32E_3} (2\pi)^4\delta(p_1+p_2+p_3-K) \left|\mathcal{M}_{EFT}(2\to 3)\right|^2 \nonumber \\
	&=\frac{\left|\mathcal{M}_{EFT}(2\to 3)\right|^2}{\mathcal{I}_{EFT}^{(2 \to 3)}} \int \frac{d^3p'_2}{(2\pi)^32E'_2}\frac{d^3p'_3}{(2\pi)^32E'_3} (2\pi)^4\delta(p'_2+p'_3-K'+p_1')  \nonumber \\
	&=\frac{\left|\mathcal{M}_{EFT}(2\to 3)\right|^2}{\mathcal{I}_{EFT}^{(2 \to 3)}} \frac{\bar{p}_{CM}(\sqrt{z_1};m_2,m_3)}{4\pi \sqrt{z_1}} \Theta(\sqrt{z_1}-m_2-m_3)\, , \label{eq:F1_EFT}
\end{align}
where $z_1=(p_1-K)^2=E_t^2+m_1^2-2E_tE_1(\bar{p}_1)$ defined for convenience, and the step function ensures that the total energy is large enough to produce on-shell particles $2$ and $3$. $\mathcal{F}^{EFT}_1(p_1)$ depends only on $\bar{p}_1$ but not its direction, so the integrals over $\vec{p}_1$ reduce to one-dimensional integrals, including the phase-space integral
\begin{align}
	\mathcal{I}_{EFT}^{(2 \to 3)}&=\int \frac{d^3p_1}{(2\pi)^32E_1} \left|\mathcal{M}_{EFT}(2\to 3)\right|^2\frac{\bar{p}_{CM}(\sqrt{z_1};m_2,m_3)}{4\pi \sqrt{z_1}} \Theta(\sqrt{z_1}-m_2-m_3) \nonumber \\
	&=\left|\mathcal{M}_{EFT}(2\to 3)\right|^2 \int_{(m_2+m_3)^2}^{(E_t-m_1)^2} dz_1\, \frac{\bar{p}_1(z_1)}{8\pi^2 E_t} \frac{\bar{p}_{CM}(\sqrt{z_1};m_2,m_3)}{4\pi \sqrt{z_1}}  \, , \label{eq:IEFT}
\end{align}
and the entanglement entropy
\begin{align}
	S_{EE}(\rho_1^{EFT})
	&=-\int \frac{d^3p_1}{(2\pi)^3 2E_1}\mathcal{F}^{EFT}_1(p_1) \log\left(\frac{\mathcal{F}^{EFT}_1(p_1)}{2E_1V}\right) \nonumber \\
	&=-\int_{(m_2+m_3)^2}^{(E_t-m_1)^2} dz_1\, \frac{\bar{p}_1(z_1)}{8\pi^2 E_t} \mathcal{F}^{EFT}_1(z_1) \log\left(\frac{\mathcal{F}^{EFT}_1(z_1)}{2E_1V}\right) \, . \label{eq:SEE_rho1_EFT}
\end{align}
On the other hand, the entanglement entropy by $\rho_{12}^{EFT}$ can be obtained easily by exchanging $1\leftrightarrow 3$ in (\ref{eq:SEE_rho1_EFT}) as follows
\begin{align}
	S_{EE}(\rho_{12}^{EFT})
	&=S_{EE}(\rho_{3}^{EFT}) \nonumber \\
	&=-\int_{(m_1+m_2)^2}^{(E_t-m_3)^2} dz_3\, \frac{\bar{p}_3(z_3)}{8\pi^2 E_t} \mathcal{F}^{EFT}_3(z_3) \log\left(\frac{\mathcal{F}^{EFT}_3(z_3)}{2E_3V}\right) \, . \label{eq:SEE_rho3_EFT} \, 
\end{align}
where $z_3=(p_3-K)^2=E_t^2+m_3^2-2E_tE_3(\bar{p}_3)$. Comparing (\ref{eq:F1_EFT}) to (\ref{eq:IEFT}), the entanglement entropies are independent of any coupling by normalization, whereas they are determined by kinematics.

\begin{figure}
	\centering
	\includegraphics[width=0.6\textwidth]{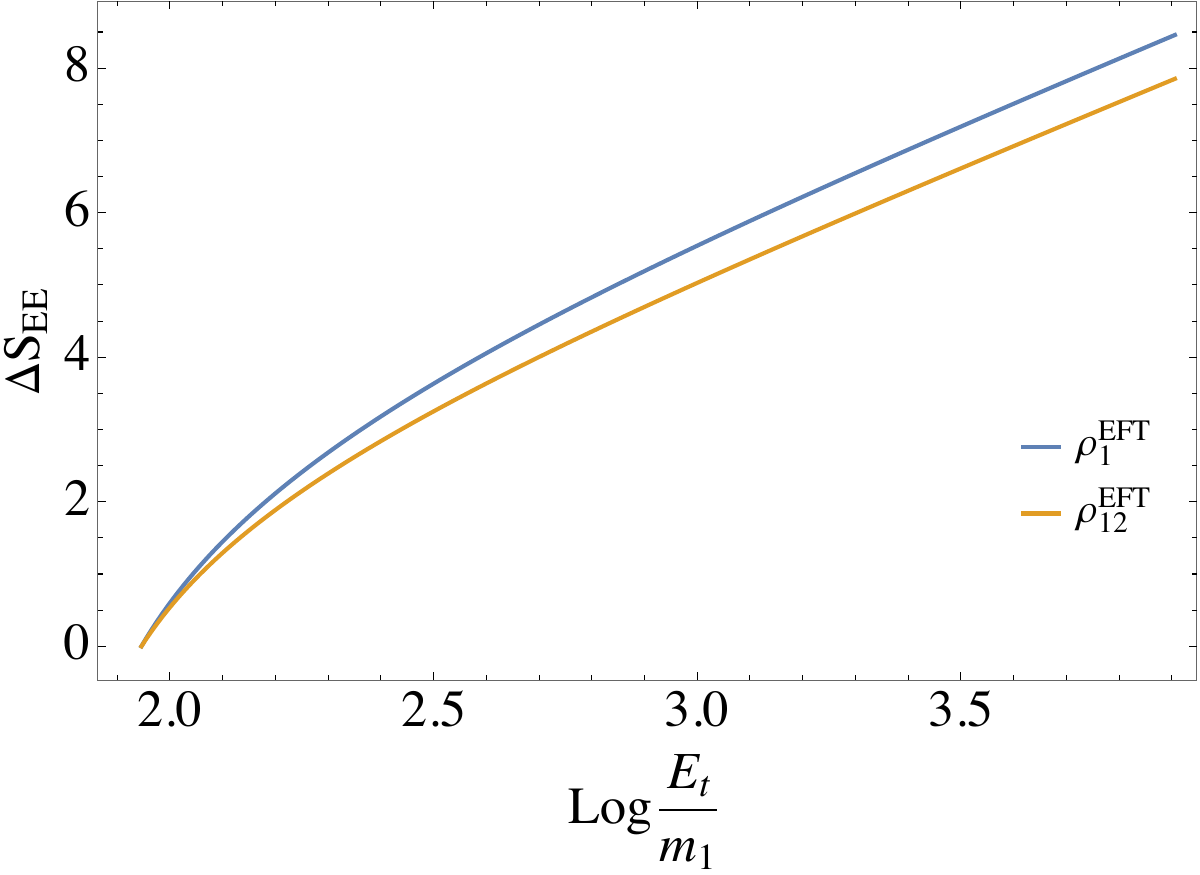}
	\caption{The changes of $S_{EE}(\rho_1^{EFT})$ and $S_{EE}(\rho_{12}^{EFT})$ with respect to the reference values at $E_{t,{\rm min}}=7m_1$, where the three masses of the final particles are $(m_1,m_2,m_3)=(1,2,3)m_1$. \label{fig:SEE_EFT}}
\end{figure}

As the entanglement entropy must involve the infinite volume $V$, there is a trick to eliminate this by considering the entropy change with respect to the value at a reference total energy $E_{t,{\rm min}}$. This utilizes the fact that the entropy has the following separation
\begin{align}
	S_{EE}&=-\int \frac{d^3p}{(2\pi)^32E}\mathcal{F}(p)\log\left(\frac{\mathcal{F}(p)}{2EV}\right) \nonumber \\
	&=-\int \frac{d^3p}{(2\pi)^32E}\mathcal{F}(p)\log\left(\frac{\mathcal{F}(p)}{2E}\right)+\log V \, ,
\end{align}
 where the normalization condition $\int \frac{d^3p}{(2\pi)^32E}\mathcal{F}(p)=1$ is applied, and the $\log V$ dependence is eliminated by subtraction $\Delta S_{EE}(E_t)=S_{EE}(E_t)-S_{EE}(E_{t,{\rm min}})$, provided that the change of volume $V$ with $E_t$ is negligible, in the sense that its change with $E_t$ is mild around the dramatic reduction of entanglement entropy due to the on-shell heavy particle. Such a trick works similarly for the entanglement entropy with probabilistic mixture (\ref{eq:SEE_1j_mixture}), where the $\log(VT/(2\pi))$ factor can also be subtracted, by assuming that both the volume $V$ and interaction time $T$ change mildly with $E_t$ compared to the dramatic reduction by intermediate heavy particles,\footnote{In practice, the effective volume $V$ and time $T$ are determined by multiple factors, including the size of wave packet and relative velocity of particles, which can change with total energy $E_t$.} and thus the changes of these geometric parameters near this energy regime can be neglected as the leading approximation. With the trick, the numerical results of (\ref{eq:SEE_rho1_EFT}) and (\ref{eq:SEE_rho3_EFT}) are shown in Figure~\ref{fig:SEE_EFT}. It is noteworthy that the entropies in the lower-energy EFT grow much faster than $\mathcal{O}(\log E_t)$, but a different tendency will appear when the intermediate heavy particle dominates the contribution.

\subsection{$S_{EE}\left(\rho^{(2\to 3)}_{12}\right)$ in the complete theory} \label{sec:numerical_12}
\begin{figure}
	\centering
	\includegraphics[width=0.7\textwidth]{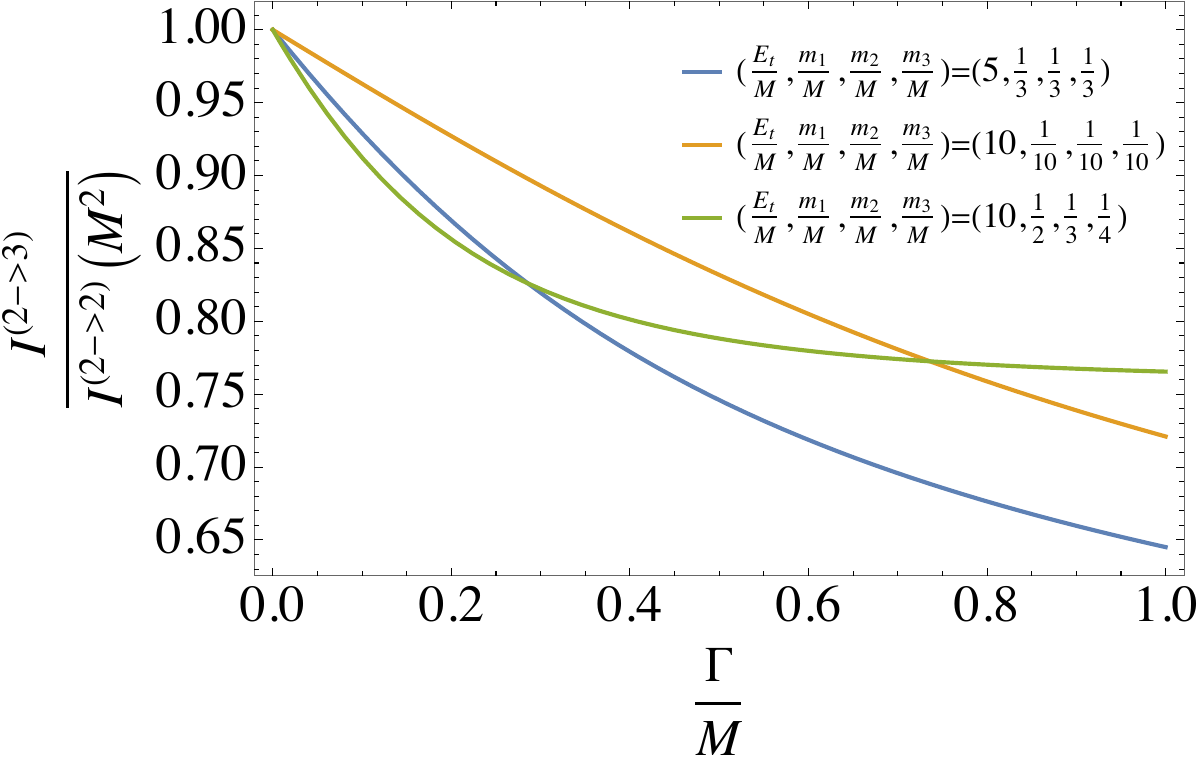}
	\caption{$\mathcal{I}^{(2\to 3)}/\mathcal{I}^{(2\to 2)}(M^2)$ versus $\Gamma/M$  with various values of $m_i$ and $E_t$. \label{fig:Ratio_I23_I22}}
\end{figure}

This corresponds to the case discussed in Section~\ref{sec:jn-1} with $(m,n)=(2,3)$ and $j=n-1$, so the reduced density matrix $\rho_{12}^{(2\to 3)}$ has the form of (\ref{eq:rho_mn_1-n-1}) with coefficients (\ref{eq:Fmn_1_n-1}): 
\begin{align}
	\mathcal{F}^{(2 \to 3)}_{12}(z) &=\frac{\pi}{\Gamma M}\mathcal{P}_{\Gamma,M}(z)\left|\mathcal{M}(2\to 2)\right|^2\frac{\mathcal{I}^{(1\to 2)}(z)}{\mathcal{I}^{(2 \to 3)}}\Theta\left(z-\left(m_1+m_2\right)^2\right) \nonumber \\
	&= 2\pi\mathcal{P}_{\Gamma,M}(z)\frac{\left|\mathcal{M}(2\to 2)\right|^2}{\mathcal{I}^{(2 \to 3)}}\frac{M\bar{p}_{CM}(\sqrt{z};m_1,m_2)}{\sqrt{z}\bar{p}_{CM}(M;m_1,m_2)}\Theta\left(z-\left(m_1+m_2\right)^2\right)\, ,
\end{align}
where the 2-body phase-space integral and its relation to $\Gamma$ are applied. The normalization ${\rm Tr}\rho_{12}^{(2\to 3)}=1$ fixes the $\mathcal{I}^{(2 \to 3)}$ as
\begin{align}
	\mathcal{I}^{(2 \to 3)}&=\left|\mathcal{M}(2\to 2)\right|^2 \int \frac{d^3q}{(2\pi)^3 2E_3} 2\pi\mathcal{P}_{\Gamma,M}(z)\frac{M\bar{p}_{CM}(\sqrt{z};m_1,m_2)}{\sqrt{z}\bar{p}_{CM}(M;m_1,m_2)}\Theta\left(z-\left(m_1+m_2\right) ^2\right) \nonumber \\
	&=\left|\mathcal{M}(2\to 2)\right|^2 \int_{(m_1+m_2)^2}^{(E_t-m_3)^2}dz \, \frac{\bar{p}_{CM}(E_t;\sqrt{z},m_3)}{4\pi E_t}  \mathcal{P}_{\Gamma,M}(z)\frac{M\bar{p}_{CM}(\sqrt{z};m_1,m_2)}{\sqrt{z}\bar{p}_{CM}(M;m_1,m_2)} \, , \label{eq:I23_num}
\end{align}
where the upper limit follows from the definition of $z(\bar{q})$ (\ref{eq:z_qbar_j_n-1}). Clearly, when $\Gamma\ll M$, the Cauchy distribution is approximately $\delta(z-M^2)$ in the way described by (\ref{eq:Cauchy_expectation}), reducing (\ref{eq:I23_num}) to a $2\to 2$ integral with the on-shell heavy particle
\begin{align}
	\mathcal{I}^{(2\to 2)}(M^2) &= \int d\Pi_2(K;q,p_3)\left|\mathcal{M}(2\to 2)\right|^2 \nonumber \\
	&=\left|\mathcal{M}(2\to 2)\right|^2 \frac{\bar{p}_{CM}(E_t;M,m_3)}{4\pi E_t} \, .
\end{align}
The ratio $\mathcal{I}^{(2\to 3)}/\mathcal{I}^{(2\to 2)}(M^2)$ is shown in Figure~\ref{fig:Ratio_I23_I22} with various $\Gamma/M \propto g^2$, confirming that the on-shell approximation for $\mathcal{I}^{(2\to 3)}$ with $\Gamma=\Gamma_{1\to 2}(M^2)$ (\ref{eq:Imn_Imn-j+1_recursion}) is valid for $\Gamma\ll M$.

\begin{figure}[t]
    \centering	\includegraphics[width=\textwidth]{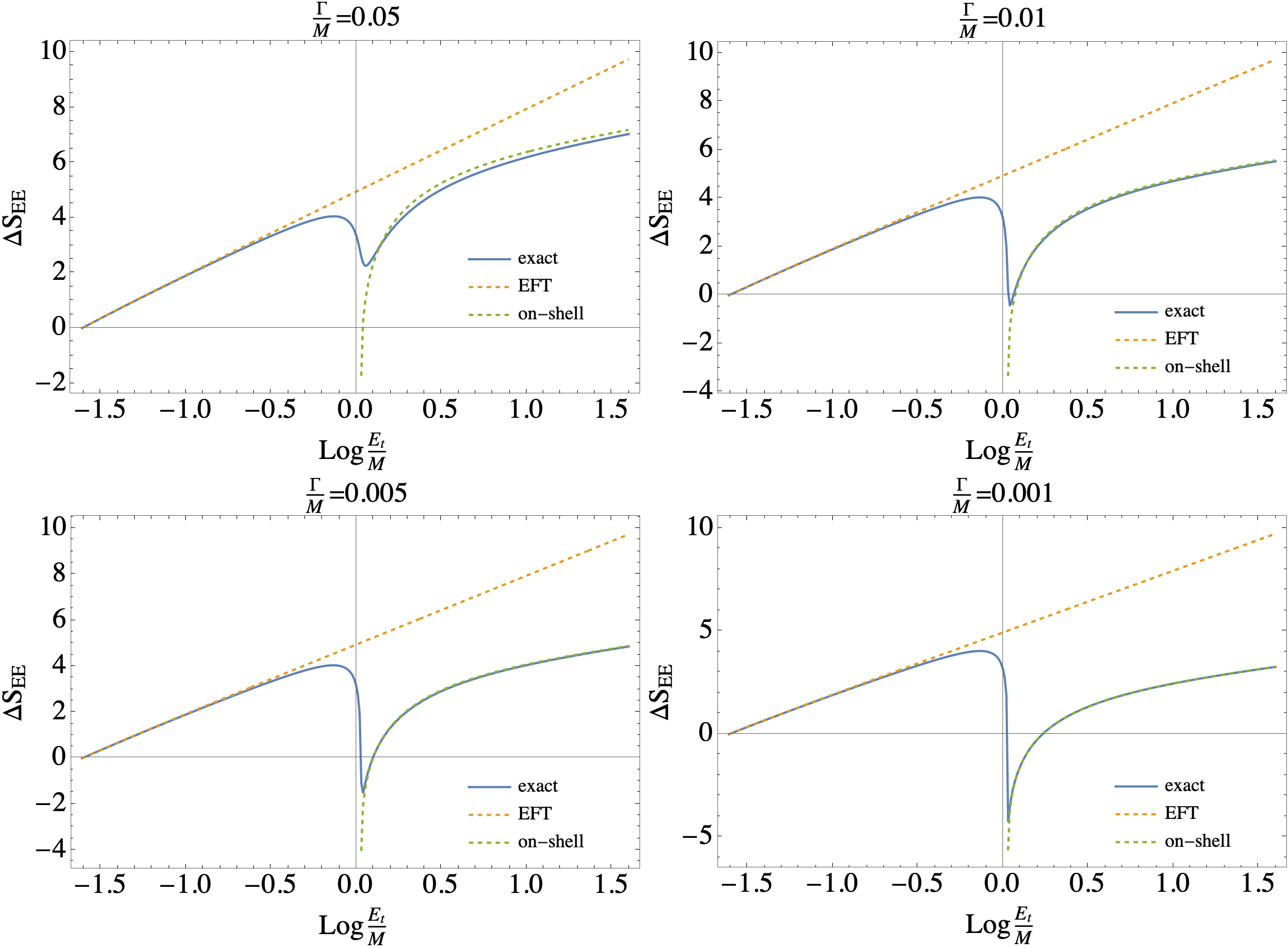}
	\caption{The change of $S_{EE}\left(\rho^{(2\to 3)}_{12}\right)$ with respect to the value at a reference total energy $E_{t,{\rm min}}=0.2M$, where the masses of the three final particles are $(m_1,m_2,m_3)=(0.01,0.02,0.03)M$, and various values of decay rate $\Gamma$ are chosen. The blue solid line is the exact numerical value of (\ref{eq:SEE_rho12_23_numerical}), whereas the yellow and green dashes lines are the EFT and on-shell approximations, (\ref{eq:SEE_rho3_EFT}) and (\ref{eq:SEE_rho_12_23_on-shell}), respectively. \label{fig:SEErho12_2to3}} 
\end{figure}

On the other hand, with (\ref{eq:SEE_1n-1_onshell_angle}), the entanglement entropy of $\rho_{12}^{(2\to 3)}$  is a one-dimensional integral
\begin{align}
	S_{EE}\left(\rho^{(2\to 3)}_{12}\right)=-\int_{(m_1+m_2)^2}^{(E_t-m_3)^2}dz \, \frac{\bar{p}_{CM}(E_t;\sqrt{z},m_3)}{8\pi^2 E_t}\mathcal{F}^{(2 \to 3)}_{12}(z) \log\left(\frac{\mathcal{F}^{(2 \to 3)}_{12}(z)}{2E_3(z)V}\right) \, , \label{eq:SEE_rho12_23_numerical}
\end{align}
and Figure~\ref{fig:SEErho12_2to3} shows its numerical value with comparison to the EFT and on-shell approximations, (\ref{eq:SEE_rho3_EFT}) and (\ref{eq:SEE_rho_12_23_on-shell}), respectively. A remarkable feature is observed that a sharp dip appears when the total energy is sufficiently large to make the heavy particle on-shell, $\log(E_t/M)\gtrsim 0$. It is noteworthy that such a dip feature is different to the peak feature commonly seen in the total cross section: when the decay rate $\Gamma$ set to be smaller, the local maximum to the left of $\log(E_t/M)=0$ remains largely unchanged, whereas the local minimum to the right becomes significantly more suppressed, even lower than the value at a reference low-energy point. Such a dip feature indicates that when the low-energy EFT starts to break down, the amount of information being propagated experiences a sharp decrease, approaching a suppressed value (by small $\Gamma$) attributed to the on-shell heavy particle. In addition, the smaller $\Gamma$ is, the more accurate the on-shell approximation (\ref{eq:SEE_rho_12_23_on-shell}) is. 

Finally, the physical meaning of the on-shell approximation $S_{EE}\left(\rho_{12}^{(2\to 3)}\right)$ can also be seen as follows. We first rearrange (\ref{eq:SEE_rho_12_23_on-shell}) into
\begin{align}
	S_{EE}\left(\rho_{12}^{(2\to 3)}\right) &\approx \log\left(\Gamma\frac{M}{E_M} \frac{\bar{p}_{CM}(E_t;M,m_3)E_M E_3V}{\pi E_t}\right) \, ,
\end{align}
where $E_M$ is the energy of the on-shell heavy particle. This can be compared with the entanglement entropy of inelastic $2\to 2$ scattering in which the final state consists of an on-shell heavy particle with mass $M$ and particle $3$ (\ref{eq:SEE22_structure}):
\begin{align}
	S_{EE}\left(\rho^{(2\to 2)}_{M}\right)&=\log\left(\frac{\bar{p}_{CM}(E_t;M,m_3)E_ME_3V}{\pi E_t T}\right)  \, ,
\end{align}
corresponding to the fact that the interaction time $T$ is replaced by the dilated lifetime in the CM frame $\frac{E_{M}}{M \Gamma}$.

\subsection{Generalization to the $2\to 4$ scattering}\label{sec:2-4_example}
\begin{figure}
	\centering
	\includegraphics[width=\textwidth]{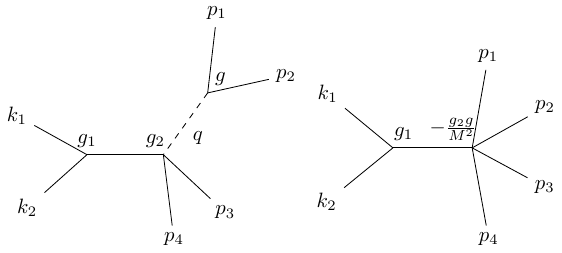}
	\caption{{\bf Left:} A simple  $2\to 4$ scattering. The solid and dashed internal lines represent particles with masses $m_\chi$ and $M$ respectively, whereas the solid external lines are light particles. {\bf Right:} An effective field theory obtained by the large $M$ expansion, and the four outgoing particles are connected by the effective vertex $-\frac{g_2g}{M^2}$. \label{fig:2-4_example}}
\end{figure}

As an example that $j\neq n-1$ discussed in Section~\ref{sec:entanglement-mton}, we verify the on-shell approximation derived with the probabilistic mixture (\ref{eq:SEE_1j_on-shell})
\begin{align}
	S_{EE}\left(\rho^{(2 \to 4)}_{12}\right) \approx \log\left(4\pi\Gamma M\right)+s\left(f^{(2\to 4)}_{12}(q) \right)\Big|_{q^2=M^2}+\log\left(\frac{VT}{2\pi}\right) \, , \label{eq:SEE_24_on_shell}
\end{align}
in a concrete model of $2\to 4$ scattering, as shown in Figure~\ref{fig:2-4_example}. Many results from the previous $2\to 3$ scattering are applicable with minor modifications.

Compared to the $2\to 3$ interacting Lagrangian (\ref{eq:L_int_2-3}), the only difference is adding one more light field $\phi_4$ at the $g_2$ vertex  
\begin{align}
	\mathcal{L}_{\rm int}=g_1\phi_A\phi_B\chi+g_2\chi\sigma\phi_3\phi_4+g\sigma\phi_1\phi_2 \, .
\end{align}
The amplitude of this process is the same as before
\begin{align}
	i \mathcal{M}{(2 \to 4)}&\approx i \mathcal{M}{(2 \to 3)} \frac{g}{(p_1+p_2)^2-M^2+i \Gamma M} \nonumber \\
	&=-i\frac{g_1g_2}{(k_1+k_2)^2-m_\chi^2+i\epsilon} \frac{g}{(p_1+p_2)^2-M^2+i \Gamma M} \, ,
\end{align}
but changing the notation of first part from $\mathcal{M}(2\to 2)$ to $\mathcal{M}(2\to 3)$. The EFT amplitude $i\mathcal{M}_{EFT}(2\to 4)$ is therefore the same as before (\ref{eq:EFT_amplitude}).

With (\ref{eq:rho_1j_mixture}), we write the reduced density matrix of the decay products (particles $1$ and $2$) into the form
\begin{align}
	\rho^{(2 \to 4)}_{12}=\int dq^2 P^{(2 \to 4)}_{12}(q^2) \sigma^{(2 \to 4)}_{12}(q^2) \, . \label{eq:rho_12_2to4_mixture}
\end{align}
The probability distribution of $q^2$ is
\begin{align}
	&P^{(2 \to 4)}_{12}(q^2) \nonumber \\
    =& \mathcal{P}_{\Gamma,M}(q^2)\frac{\mathcal{I}^{(1\to 2)}(q^2) \mathcal{I}^{(2 \to 3)}(q^2)}{2\Gamma M\mathcal{I}^{(2 \to 4)}}\Theta\left(q^2-\left(m_1+m_2\right)^2\right) \nonumber \\
	=&\mathcal{P}_{\Gamma,M}(q^2)\frac{M\bar{p}_{CM}(\sqrt{q^2};m_1,m_2)}{\sqrt{q^2}\bar{p}_{CM}(M;m_1,m_2)}\frac{\left|\mathcal{M}(2\to 3)\right|^2}{\mathcal{I}^{(2 \to 4)}}\int d\Pi_3(K;q,p_3,p_4)\Theta\left(q^2-\left(m_1+m_2\right)^2\right)
	\, \label{eq:P_12_2to4_mixture} \, ,
\end{align}
where the 3-body phase-space integral is evaluated
\begin{align}
	\int d\Pi_3(K;q,p_3,p_4)
	&=\int_{(m_3+m_4)^2}^{\left(E_t-\sqrt{q^2}\right)^2} dz\, \frac{\bar{q}(z)}{8\pi^2 E_t} \frac{\bar{p}_{CM}(\sqrt{z};m_3,m_4)}{4\pi \sqrt{z}} \, , \label{eq:3_body_q_3_4}
\end{align}
by applying the previous result (\ref{eq:IEFT}), with the replacements $(p_1,p_2,p_3)\to (q,p_3,p_4)$ and $z=(K-q)^2=E_t^2+q^2-2E_t\sqrt{q^2+\bar{q}^2}$. Clearly, the upper and lower limits of (\ref{eq:3_body_q_3_4}) also fix the upper limit of $q^2\leq (E_t-m_3-m_4)^2$, and the normalization $\int dq^2P^{(2 \to 4)}_{12}(q^2)=1$ in (\ref{eq:P_12_2to4_mixture}) fixes the denominator
\begin{align}
	\mathcal{I}^{(2\to 4)}=&\left|\mathcal{M}(2\to 3)\right|^2\int_{(m_1+m_2)^2}^{(E_t-m_3-m_4)^2} dq^2\mathcal{P}_{\Gamma,M}(q^2) \frac{M\bar{p}_{CM}(\sqrt{q^2};m_1,m_2)}{\sqrt{q^2}\bar{p}_{CM}(M;m_1,m_2)}\nonumber \\
    &\times\int_{(m_3+m_4)^2}^{\left(E_t-\sqrt{q^2}\right)^2} dz\, \frac{\bar{q}(z)}{8\pi^2 E_t} \frac{\bar{p}_{CM}(\sqrt{z};m_3,m_4)}{4\pi \sqrt{z}}  \, . \label{eq:I24_integral}
\end{align}
On the other hand, with (\ref{eq:sigma_1j_mixture}), the state's part of reduced density matrix is 
\begin{align}
	\sigma^{(2 \to 4)}_{12}(q^2) &=\int d\Pi_{3}(K;q,p_3,p_4)\frac{\left|\mathcal{M}(2\to 3)\right|^2}{\mathcal{I}^{(2\to 3)}(q^2)}  |\tilde{\psi}^{(2\to 4)}_{12}(q)\rangle \langle \tilde{\psi}^{(2\to 4)}_{12}(q)| \nonumber \\
	&=\frac{1}{\int d\Pi_3(K;q,p_3,p_4)} \int \frac{d^3q}{(2\pi)^3 2q^0}\frac{\bar{p}_{CM}(\sqrt{z};m_3,m_4)}{4\pi \sqrt{z}}|\tilde{\psi}^{(2\to 4)}_{12}(q)\rangle \langle \tilde{\psi}^{(2\to 4)}_{12}(q)| \, ,
\end{align}
and the corresponding coefficient is
\begin{align}
	f^{(2\to 4)}_{12}(q)=\frac{1}{\int d\Pi_3(K;q,p_3,p_4)} \frac{\bar{p}_{CM}(\sqrt{z};m_3,m_4)}{4\pi \sqrt{z}} \, .
\end{align}
With these and (\ref{eq:SEE_1j_mixture}), the entanglement entropy is 
\begin{align}
	S_{EE}\left(\rho^{(2 \to 4)}_{12}\right)
	&=h\left(P^{(2 \to 4)}_{12}(q^2)\right) +\int dq^2 P^{(2 \to 4)}_{12}(q^2) s\left(f^{(2\to 4)}_{12}(q) \right)+\log\left(\frac{VT}{2\pi}\right) \, , \label{eq:SEE_rho12_24_numerical}
\end{align}
where the differential entropies are defined in (\ref{eq:h_1j_mixture}) and (\ref{eq:s_1j_state}), respectively, and the latter can be simplified as the integrand involving $\vec{q}$ is spherically symmetric
\begin{align}
	s\left(f^{(2\to 4)}_{12}(q) \right)&=-\int d\bar{q}\frac{4\pi \bar{q}^2}{(2\pi)^3 2 q^0}f^{(2\to 4)}_{12}(q)\log\left(f^{(2\to 4)}_{12}(q)\right) \nonumber \\
	&=-\int_{(m_3+m_4)^2}^{\left(E_t-\sqrt{q^2}\right)^2} dz\, \frac{\bar{q}(z)}{8\pi^2 E_t} f^{(2\to 4)}_{12}(z)\log\left(f^{(2\to 4)}_{12}(z)\right)\, .
\end{align} 
For the same quantities defined in the EFT, we can simply replace the Cauchy distribution $\mathcal{P}_{\Gamma,M}(q^2)$ with a uniform distribution
\begin{align}
	\mathcal{P}_{EFT}=\frac{\Gamma M}{\pi M^4} \, , \label{eq:24_EFT}
\end{align}
corresponding to the large $M$ expansion, and the $\Gamma$ dependence is eventually eliminated by the normalization of $P^{(2\to 4)}_{12}(q^2)$.

\begin{figure}
	\centering
	\includegraphics[width=0.6\textwidth]{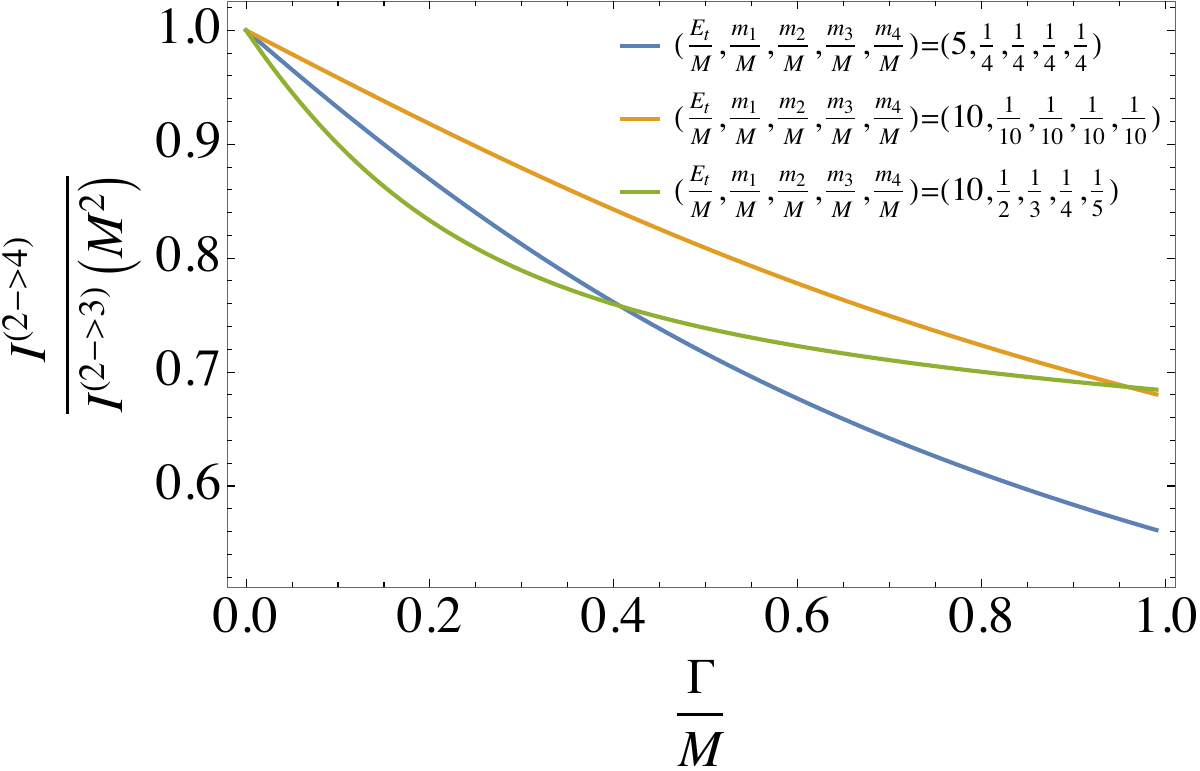}
	\caption{$\mathcal{I}^{(2\to 4)}/\mathcal{I}^{(2\to 3)}(M^2)$ versus $\Gamma/M$  with various values of $m_i$ and $E_t$. \label{fig:Ratio_I24_I23}}
\end{figure}

\begin{figure}
	\centering
	\includegraphics[width=\textwidth]{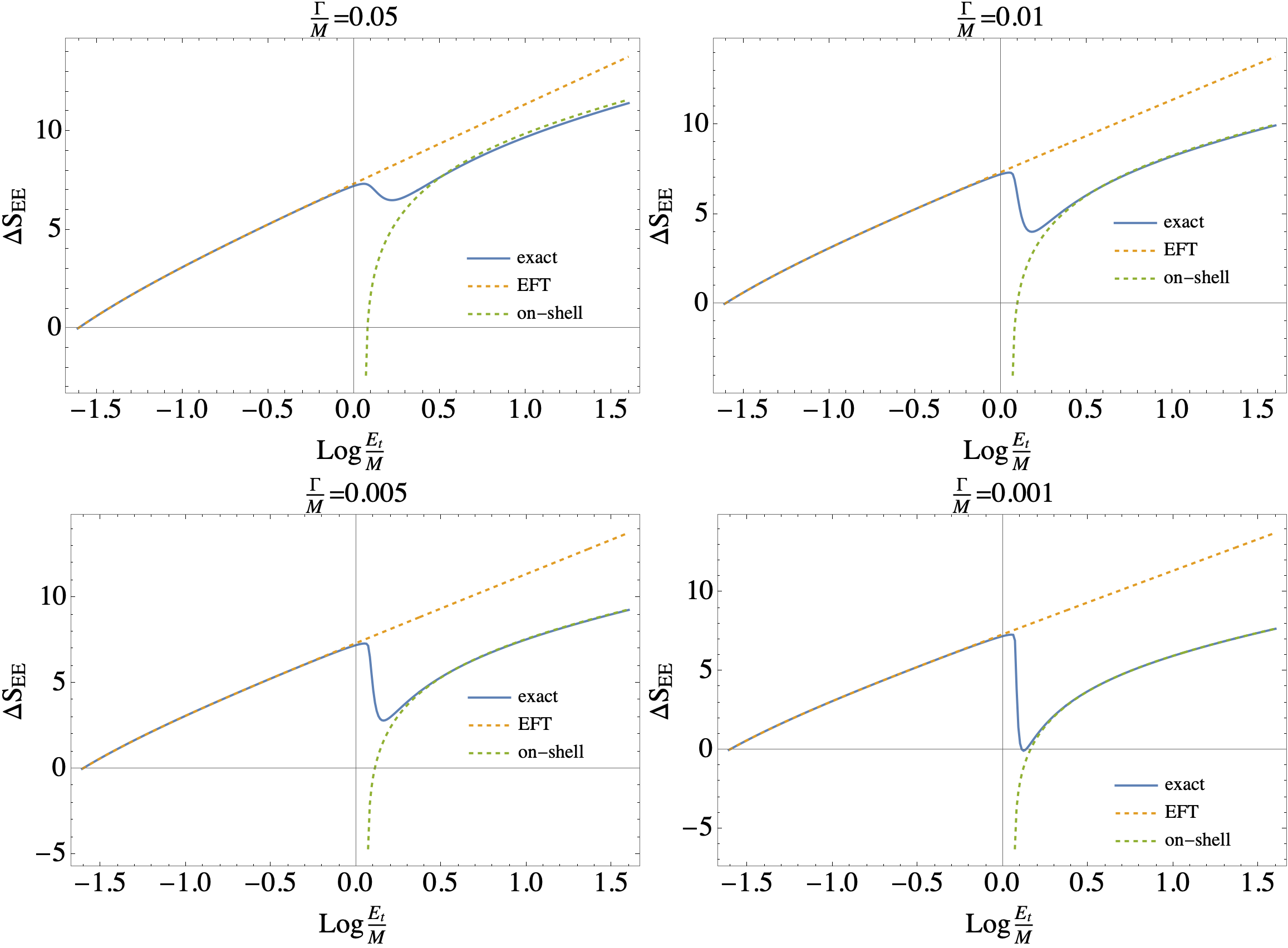}
	\caption{The change of $S_{EE}\left(\rho^{(2\to 4)}_{12}\right)$ with respect to the value at a reference total energy $E_{t,{\rm min}}=0.2M$, where the masses of the four final particles are $(m_1,m_2,m_3,m_4)=(0.01,0.02,0.03,0.04)M$, and various values of decay rate $\Gamma$ are chosen. The blue solid line is the exact numerical value of (\ref{eq:SEE_rho12_24_numerical}), whereas the yellow and green dashes lines are the EFT and on-shell approximations, (\ref{eq:24_EFT}) and (\ref{eq:SEE_24_on_shell}), respectively. \label{fig:SEErho12_2to4}} 
\end{figure}

As shown in Figure~\ref{fig:Ratio_I24_I23}, we first verify that the ratio $\mathcal{I}^{(2\to 4)}/\mathcal{I}^{(2\to 3)}(M^2)\approx 1$ (\ref{eq:Imn_Imn-j+1_recursion}) is valid for $\Gamma\ll M$ and $\Gamma=\Gamma_{1\to2}(M^2)$, as the on-shell approximation of entanglement entropy relies on this relation. The entanglement entropy of the $2\to 4$ scattering is then computed numerically, compared with the low-energy EFT (\ref{eq:24_EFT}) and the on-shell approximation (\ref{eq:SEE_24_on_shell}), as shown in Figure~\ref{fig:SEErho12_2to4}. Similar to the previous $2\to 3$ scattering, the sharp dip located at $\log(E_t/M)\gtrsim 0$ is observed, and the smaller the $\Gamma$, the more accurate the on-shell approximation. The results of $2\to 4$ scattering confirm that such a sharp dip is attributed to the probabilistic mixture associated to the Cauchy distribution $\mathcal{P}_{\Gamma,M}(q^2)$, as derived in Section~\ref{sec:structure_decay_onshell}, and such features are expected to appear for general $m\to n$ inelastic scattering with intermediate heavy particles.

\section{Conclusion}\label{sec:conclusion}
We have explored the information propagated by an intermediate heavy particle in general $m\to n$ inelastic scattering, in which the amount of information is quantified by the entanglement entropy between its decay products and other particles unrelated to the decay. When the kinematics allows the heavy particle to be on-shell, it is expected from the form of its propagator that the entanglement is dominated by this limit in the phase space. We have proved that such an intuition is correct, by using the properties of Cauchy distribution, and in fact the on-shell contribution is suppressed universally by the small decay rate $\Gamma$. Such an entanglement suppression originates from the probabilistic mixture by the sharp-peak nature of heavy-field propagator around the on-shell limit, and the small uncertainty by the decay rate gives the suppressed entropy with the form of $\log(4\pi\Gamma M c)$.

As the low-energy EFT does not depend on the small decay rate, it is expected that the entanglement at low energy is unsuppressed, implying that the entanglement entropy experiences a dramatic reduction when the EFT starts to break down at $E_t\approx M$ for approaching the on-shell limit. We have shown that this reasoning is also correct in concrete models of $2\to 3$ and $2\to 4$ scatterings, in which the more suppressed the entanglement, the more accurate the on-shell approximation. 

Experimentally, the entanglement features, seen in the marginalization of the phase-space distribution $\propto\left|\mathcal{M}(m\to n)\right|^2(p_1,\dots,p_n)$, can in principle be measured by statistically analyzing the particle detector's data. For example, when the decay products entangle with a particle $n$, the reduced density matrix $\rho_{n}^{(m\to n)}$ (\ref{eq:rho_n_mton}) is constructed by marginalizing $(p_1,\dots,p_{n-1})$, giving the statistical value of entanglement entropy. This might be a guide for the development of new quantum-information observables, such as a reduced amount of the Bell violation.

Theoretically, such clear features might also have the potential to analyze the breakdown of EFT, such as developing constraints for EFT based on the expected suppression of entanglement. We leave it for future study.

Technically, we have also demonstrated an example that relates quantum-information quantities to the $S$-matrix pole structure, leading to features beyond the area law in $2 \to 2$ scattering. Some relations with $S$-matrix bootstrap in $2\to2$ scattering have also been discussed in the literature \cite{Bose:2020shm,Sinha:2022crx}, and our results indicate the value of going beyond the $2\to 2$ case.

\section*{Acknowledgments}
CMS thanks Qi Chen for helpful discussion, and Zhong-Zhi Xianyu for valuable discussions and comments on the results. CMS is supported by the Shuimu Tsinghua scholar program, NSFC under Grant No. 12275146, the National Key R\&D Program of China (2021YFC2203100), and the Dushi Program of Tsinghua University. YW and XZ are supported in part by the RGC Research Fellow Grant RFS2425-6S02 from the RGC of Hong Kong.

\appendix
\section{The factor of $4$ in the entanglement entropy of decay}\label{sec:factor_4}
We take the entanglement entropy by $1\to 2$ decay obtained in \cite{Lello:2013bva} as an example to demonstrate the factor-of-4 subtlety of Cauchy distribution, as stated in (\ref{eq:Cauchy_entropy}) and (\ref{eq:wrong_Cauchy_entropy}).
The entropy obtained by the Wigner-Weisskopf method is 
\begin{equation}
S_{EE}^{(1\to 2)} = -\sum_{p} n(p) \log n(p)\, ,\label{entropy by sum over spasespace}
\end{equation}
where the number density of states is
\begin{equation}
n(p) = \frac{2\pi^2 M}{V \bar{p} E_1 E_2}  \frac{1}{\pi} \frac{\Gamma/2}{(E_1 + E_2 - M)^2 + \Gamma^2/4}\, .\label{phasespace density}
\end{equation}
The peak of $n(p)$ occurs at the CM momentum of the two final particles
\begin{equation}
\bar{p}^* = \bar{p}_{CM}(M;m_1,m_2)\, ,\label{momentum in CoM}
\end{equation} 
As $n(p)$ is spherically symmetric and satisfies the normalization $\sum_p n(p) = 1$, the authors in \cite{Lello:2013bva} applied the direct sharp-peak approximation (\ref{eq:wrong_Cauchy_entropy}) to approximate the entanglement entropy as $-\log n({p}^*)$, corresponding to 
\begin{equation}
   S_{EE}^{\rm sharp}= -\log\left[\frac{4\pi M}{V \bar{p}^* E_1(\bar{p}^*) E_2(\bar{p}^*) \Gamma}\right] \, .\label{decay entropy in the paper}
\end{equation}
However, this approach is problematic because the logarithmic factor $\log n(p)$ also has a peak near $\bar{p} = \bar{p}^*$. Given that the dominant contribution to the entropy originates from the region around $\bar{p} = \bar{p}^*$, this region requires more careful treatment. Nevertheless, the direct sharp-peak approximation is still helpful for obtaining reasonable estimations.

Now, we can consider the generalized case that the entropy integral has the following form
\begin{align}
&-\int d {x} f'(x) \frac{1}{\pi} \frac{\gamma}{f^2 + \gamma^2} \log\left(f'(x) \frac{1}{\pi} \frac{\gamma}{f^2 + \gamma^2}\right) \nonumber \\
=& -\int d{x} f'(x) \frac{1}{\pi} \frac{\gamma}{f^2 + \gamma^2} \left[\log\left(\frac{1}{\pi} \frac{\gamma}{f^2 + \gamma^2}\right) + \log f'(x)\right] \nonumber \\
\approx &-\log\left(\frac{1}{4\pi \gamma}\right) - \int d{f(x)} \frac{1}{\pi} \frac{\gamma}{f^2(x) + \gamma^2} \log f'(x) \, . 
\label{generalized Cauchy integral}
\end{align}
where  $f(x)$ is a monotonic function whose range includes zero, and (\ref{eq:Cauchy_entropy}) is applied as the primary contribution comes from the neighborhood of $f(x) = 0$:
\begin{align}
&-\int_{x({f_{\text{min}}})}^{x({f_{\text{max}})}} d{x} f'(x) \frac{1}{\pi} \frac{\gamma}{f^2 + \gamma^2} \log\left(\frac{1}{\pi} \frac{\gamma}{f^2 + \gamma^2}\right) \nonumber \\
\approx &-\int_{-\infty}^{\infty} d{x} f'(x) \frac{1}{\pi} \frac{\gamma}{f^2 + \gamma^2} \log\left(\frac{1}{\pi} \frac{\gamma}{f^2 + \gamma^2}\right) \nonumber \\
=& -\log\left(\frac{1}{4\pi \gamma}\right) \, .
\end{align}
For the second term of (\ref{generalized Cauchy integral}), the sharp-peak approximation is valid if $\log f'(x)$ varies much more slowly compared to $\frac{1}{\pi} \frac{\gamma}{f^2(x) + \gamma^2}$ in the neighborhood of $f(x) = 0$.

For the number density (\ref{phasespace density}), we have $\gamma=\frac{\Gamma}{2}$ and
\begin{equation}
f(\bar{p}) = E_1 + E_2 - M, \quad f'(\bar{p}) = \frac{\bar{p} M}{E_1 E_2} \, .
\end{equation}
Since the integration measure is given by:
\begin{equation}
V \int_0^\infty \frac{4\pi \bar{p}^2 d{\bar{p}}}{(2\pi)^3} \, ,
\label{measure for phasespace integral}\end{equation}
we should properly normalize the expression, by introducing a factor of $\frac{2\pi^2 M}{V \bar{p} E_1 E_2}$ both inside and outside the logarithm, ensuring that:
\begin{equation}
V \int_0^\infty \frac{4\pi \bar{p}^2 d{\bar{p}} }{(2\pi)^3} \frac{2\pi^2 M}{V \bar{p} E_1 E_2} = \int d{\bar{p}} f'(\bar{p}) \, .
\end{equation}
So the entanglement entropy (\ref{entropy by sum over spasespace}) is rewritten into the integral 
\begin{align}
S_{EE}^{(1\to 2)} &= -V \int_0^\infty \frac{4\pi \bar{p}^2 d{\bar{p}}}{(2\pi)^3} \frac{2\pi^2 M}{V \bar{p} E_1 E_2} \frac{1}{\pi} \frac{\Gamma/2}{(E_1 + E_2 - M)^2 + \Gamma^2/4} \nonumber \\
&\quad \times \log\left[\frac{2\pi^2 M}{V \bar{p} E_1 E_2} \frac{1}{\pi} \frac{\Gamma/2}{(E_1 + E_2 - M)^2 + \Gamma^2/4}\right]\nonumber \\
&= -\int_0^\infty d{\bar{p}} f'(\bar{p}) \frac{1}{\pi} \frac{\Gamma/2}{f^2 + \Gamma^2/4} \left\{\log\left[\frac{1}{\pi} \frac{\Gamma/2}{f^2 + \Gamma^2/4}\right] + \log\left[\frac{2\pi^2 M}{V \bar{p} E_1 E_2}\right]\right\} \, ,
\end{align}
which can be approximated with (\ref{generalized Cauchy integral}) which relies on the complex analysis done in Section~\ref{sec:on-shell_Cauchy}
\begin{align}
S_{EE}^{\rm complex}& \approx -\log\left(\frac{1}{2\pi \Gamma}\right) -\int_0^\infty d\bar{p}\, f'(\bar{p}) \frac{1}{\pi} \frac{\Gamma/2}{f^2 + \Gamma^2/4} \log\left[\frac{2\pi^2M}{V \bar{p} E_1E_2}\right] \nonumber \\
&\approx -\log\left(\frac{1}{2\pi \Gamma}\right) - \log\left[\frac{2\pi^2 M}{V \bar{p}^* E_1(\bar{p}^*) E_2(\bar{p}^*)}\right]\nonumber \\
&= -\log\left[\frac{\pi M}{V \bar{p}^* E_1(\bar{p}^*) E_2(\bar{p}^*) \Gamma}\right] \, , 
\label{particle decay entropy approximation}
\end{align}
with the fact that there is no peak is the second term $\log\left[\frac{2\pi^2M}{V \bar{p} E_1E_2}\right]$ near $\bar{p}=\bar{p}^*$.

\begin{figure}[h]
    \centering
    \includegraphics[width=\textwidth]{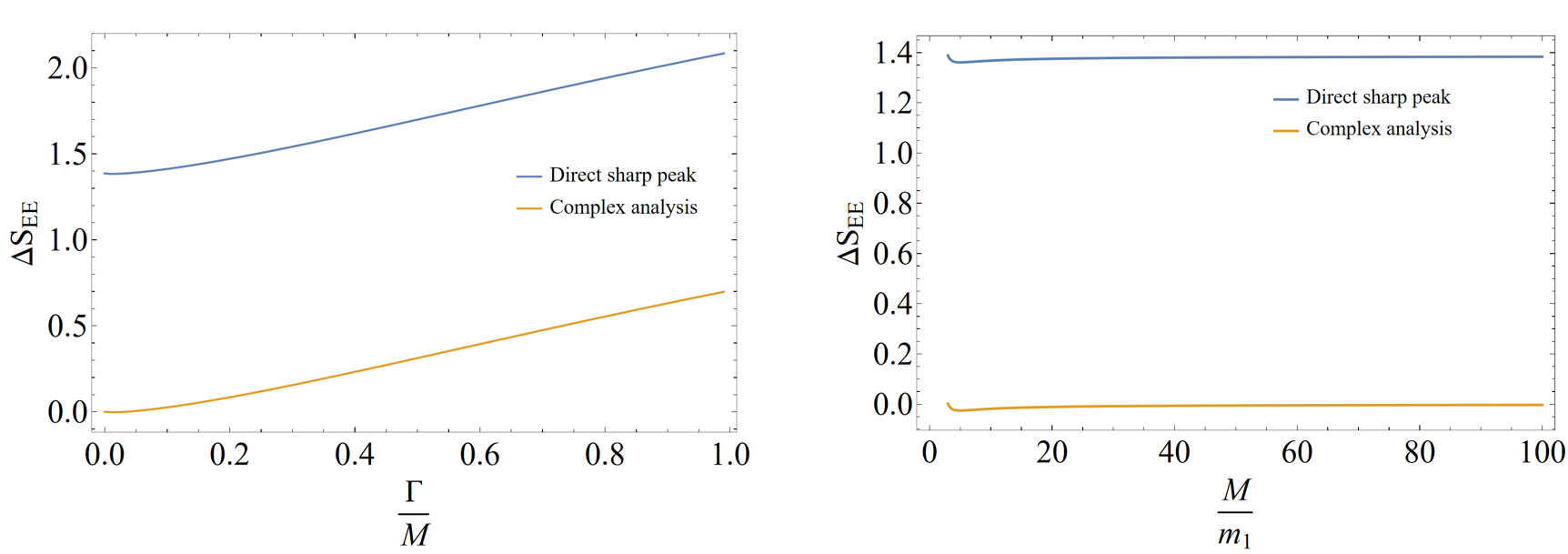}
    \caption{The differences of entanglement entropy with various parameters. The blue line is the difference $S_{EE}^{(1\to2)}-S_{EE}^{\rm sharp}$ with (\ref{decay entropy in the paper}), and the yellow line is the difference $S_{EE}^{(1\to2)}-S_{EE}^{\rm complex}$ with (\ref{particle decay entropy approximation}). {\bf Left:} Describing how the differences change when $\Gamma/M$ grows, where the parameters $(M, m_1, m_2)=(1, 0.1, 0.1)M$. {\bf Right:} Describing how the differences change when $M/m_1$ grows, where the parameters $(m_1, m_2, \Gamma)=(1, 1, 0.1)m_1$.}
    \label{fig:entropy for particle decay}
\end{figure}

To justify the discrepancy between (\ref{decay entropy in the paper}) and (\ref{particle decay entropy approximation}), we compute the decay entropy numerically.  As shown in Figure~\ref{fig:entropy for particle decay}, the entropy difference $S_{EE}^{(1\to2)}-S_{EE}^{\rm complex}$ works much better than $S_{EE}^{(1\to2)}-S_{EE}^{\rm sharp}$, whereas the latter clearly has a non-zero shift.

\bibliographystyle{utphys.bst}
\bibliography{reference_entropy_S_matrix} 

\end{document}